\pdfoutput=1
\documentclass[11pt]{article}

\usepackage{ACL2023}
\usepackage{times}
\usepackage{latexsym}

\usepackage[T1]{fontenc}
\usepackage[utf8]{inputenc}

\usepackage{microtype}

\usepackage{graphicx}
\usepackage{multicol}
\usepackage{multirow}
\usepackage{amsmath}
\usepackage{amssymb}
\usepackage{bbding}
\usepackage{subfigure}
\usepackage{colortbl}
\usepackage{pifont}
\usepackage{bbm}
\usepackage{makecell}
\usepackage{bbding}
\usepackage{mathrsfs}
\usepackage{bm}
\usepackage{CJKutf8}
\usepackage{booktabs}
\usepackage{tabu}
\usepackage{fontawesome}

\definecolor{ClosedColor}{HTML}{D8EEF2}  % 你也可以换成 cyan!15 或其他
\definecolor{OpenColor}{HTML}{FDEBDD}    % 你也可以换成 red!15 或其他
\definecolor{HeaderColor}{gray}{.85}
\usepackage{colortbl}
\arrayrulecolor{black}    % 恢复默认的表格线颜色

\usepackage[linesnumbered,ruled,vlined]{algorithm2e}
\usepackage{listings} % 集成"输入即所得"
\usepackage[most]{tcolorbox} % prompt支持
\tcbuselibrary{listings} % 加载 tcolorbox 的 listings 支持
\tcbuselibrary{breakable} % 跨页显示

\newcommand{\boxfigure}[4]{% Use % to avoid spurious spaces at end of lines
    \begin{figure}[htbp] % Use htbp for better float placement than just h
    \centering
    \begin{tcolorbox}[
        width=1.0\textwidth,
        colback=blue!2!white,
        colframe=blue!20!gray,
        arc=2mm,
        boxrule=0.5pt,
        title={\texttt{~~~~~~~~~~}}, % Argument 1: Title
        enhanced,
        fonttitle=\normalsize\bfseries,
        coltitle=black,
        attach boxed title to top left={xshift=0.3cm, yshift=-1.5mm},
        boxed title style={colback=blue!10!white, colframe=blue!20!gray, arc=1mm, boxrule=0pt},
        listing only, % Assumes content is always code
        listing options={
            breaklines=true,
            basicstyle=\ttfamily\footnotesize,
            keywordstyle=\color{blue!70!black},
            commentstyle=\color{green!60!black},
            showstringspaces=false
        },
    ]
    #2 % Argument 2: Content
    \end{tcolorbox}
    \caption{#3} % Argument 3: Caption
    \label{#4} % Argument 4: Label
    \end{figure}% Use % to avoid spurious spaces
}

\newcommand{\bench}{LiveRepoReflection}
\newcommand{\instruction}{RepoReflection-Instruct}
\newcommand{\coder}{RepoReflectionCoder}

\newcommand{\new}{full-file code generation}
\newcommand{\fix}{patch-based incremental edits}

\NewDocumentCommand\emojipaper{}{
$\vcenter{\hbox{\includegraphics[height=2em]{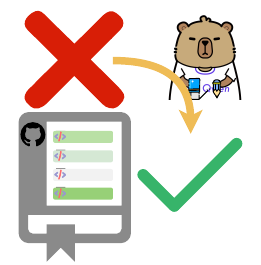}}}$
}

\title{\emojipaper{}Turning the Tide: Repository-based Code Reflection}

\author{
Wei Zhang\textsuperscript{\textrm{1}},~Jian Yang\textsuperscript{\textrm{1}\thanks{~~Corresponding author}},~Jiaxi Yang\textsuperscript{\textrm{2}},~Ya Wang,~Zhoujun Li\textsuperscript{\textrm{1}}, \\ \bf~Zeyu Cui,~Binyuan Hui,~Junyang Lin \\
\textsuperscript{\textrm{1}}CCSE, Beihang University \\
\textsuperscript{\textrm{2}}Shenzhen Institute of Advanced Technology, Chinese Academy of Sciences \\
\texttt{zwpride@buaa.edu.cn} \\
\href{https://LiveRepoReflection.github.io}{{\tt https://LiveRepoReflection.github.io}}
}

\begin{document}
\begin{CJK*}{UTF8}{gbsn}
\maketitle
\begin{abstract}
Code large language models (LLMs) enhance programming by understanding and generating code across languages, offering intelligent feedback, bug detection, and code updates through reflection, improving development efficiency and accessibility. While benchmarks (e.g. HumanEval/LiveCodeBench) evaluate code generation and real-world relevance, previous works ignore the scenario of modifying code in repositories. Considering challenges remaining in improving reflection capabilities and avoiding data contamination in dynamic benchmarks, we introduce \bench{}, a challenging benchmark for evaluating code understanding and generation in multi-file repository contexts, featuring 1,888 rigorously filtered test cases across $6$ programming languages to ensure diversity, correctness, and high difficulty. Further, we create \instruction{}, a large-scale, quality-filtered instruction-tuning dataset derived from diverse sources, used to train \coder{} through a two-turn dialogue process involving code generation and error-driven repair. The leaderboard evaluates over 40 LLMs to reflect the model performance of repository-based code reflection.
\end{abstract}

\section{Introduction}
Code large language models (LLMs)~\citep{qwen25coder,qwen3,llama,claude37,gpt4} represent a significant advancement in comprehending and producing code across numerous programming languages. Fueled by extensive training on massive code repositories, LLMs empower developers by offering intelligent feedback, identifying potential bugs, and updating code snippets from human instructions. Code reflection refers to the ability of LLMs to examine and modify their previous responses. Using reflection, LLMs can streamline the development process, boost efficiency, and make programming more accessible to a wider number of developers. 

\begin{figure}[t]
\begin{center}
    \includegraphics[width=0.8\columnwidth]{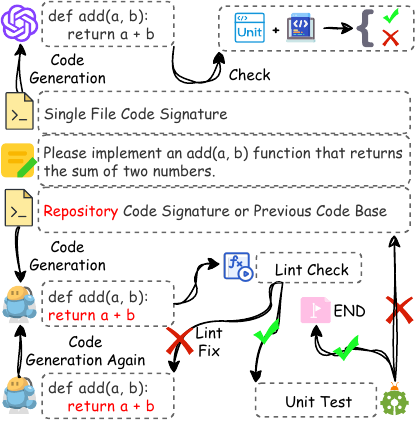}
    \caption{Comparison between general code generation tasks and \bench{} tasks.}
    \label{fig:intro}
    \vspace{-25pt}
\end{center}
\end{figure}
Most previous works primarily focus on generating and evaluating functionally correct code from human instructions illustrated in \autoref{fig:intro}, such as HumanEval/MBPP~\citep{codex,mbpp} and MultiPL-E~\citep{multiple}. More recent LLMs (e.g. Claude3.7~\citep{claude37} and Qwen3~\citep{qwen3}) and code benchmarks (e.g.  LiveCodeBench~\citep{livecodebench}, McEval~\citep{mceval}, and BigCodeBench~\citep{bigcodebench}) aim to increase the difficulty, complexity, and real-world relevance by introducing more libraries, algorithms, and programming languages. The recently proposed benchmark Aider-polyglot~\citep{aiderpolyglotCodeEditing} covering $6$ languages measures the ability of LLMs to apply code changes to source files without human intervention, reflecting a more realistic development workflow. However, the Aider benchmark is limited by a repository of programming exercises, where the repositories have been included in the pre-training data. There is still a lack of knowledge on how to improve the repository-based code reflection capability of LLMs and build dynamic benchmarks to measure the actual LLM capabilities to avoid data hacking.

To explore the LLM capability of repository-based code reflection, we propose an automatic creation pipeline to dynamically update the large-scale instruction corpus and the evaluation benchmark to avoid data hacking. This paper introduces \bench{}, a challenging benchmark for evaluating code understanding and generation in repository-style multi-file contexts, drawing inspiration from the Aider benchmarks and Exercism problems. To ensure data consistency and facilitate realistic evaluations, \bench{} comprises coding problems organized with a defined repository structure, including problem definitions, reference answers, code signatures, unit tests, and environment support files. Unlike the potentially redundant data from Exercism, \bench{} employs an optimized and streamlined file structure. Furthermore, the paper details the construction of \instruction{}, a large-scale instruction tuning dataset derived from diverse sources and filtered for quality, which is used to train \coder{}, a code-focused large language model. The data generation process for \instruction{} involves a sophisticated multi-turn dialogue simulation encompassing code generation, error-driven repair, and style standardization, aiming to enhance the model's ability to handle complex coding tasks and iterative interactions.

\begin{figure*}[t!]
\begin{center}
    \includegraphics[width=1.0\textwidth]{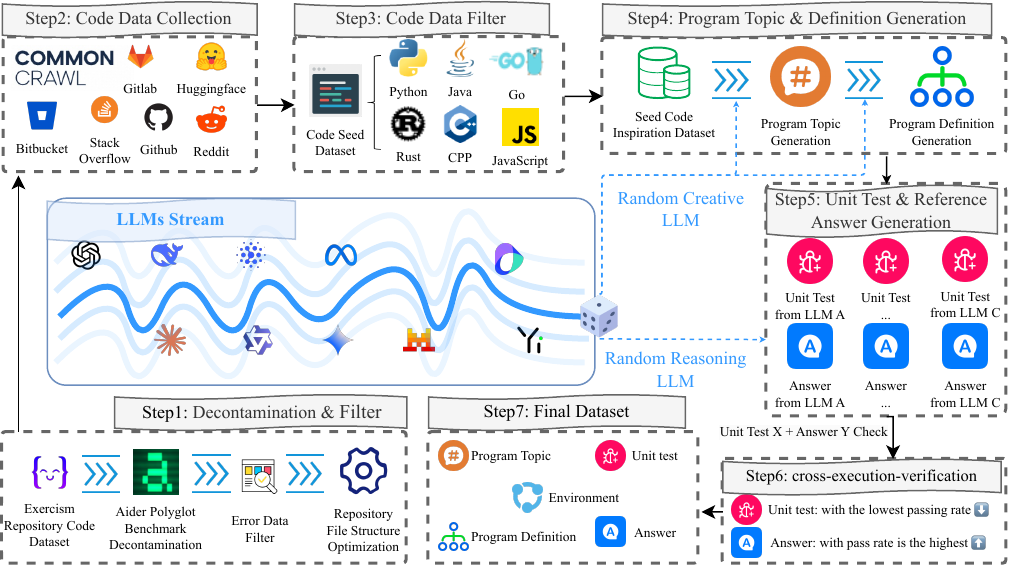}
    \caption{Overview of Construction Pipeline. We use this pipeline to generate polyglot repository code data of \bench{} and instruction corpus of \instruction{} following our designed repository code file structure. 1) Pull Exercism repos, dedupe against other benchmarks, clean out broken samples, and optimize the repository file structure. 2) Collect code snippets from multiple public source like GitHub, Hugging Face and Stack Overflow.  3)  Filter data by programming languages. 4) Use a randomized LLM stream to pick a ``creative'' LLM for generating program topics and definitions from the seed data. 5) Use several ``reasoning'' LLMs to produce unit tests and reference solutions. 6) Cross execute for verifying every unit test-solution pair, drop anomalies, then retain the test with the lowest pass rate and the solution with the highest and manual review. 7) Package everything into the final repository structure.}
    \label{fig:framework}
    \vspace{-20pt}
\end{center}
\end{figure*}

The contributions are summarized as follows: 
\begin{itemize}
    \item We propose a high-quality, high-difficulty benchmark for evaluating code reflection of LLMs, featuring 1,888 rigorously filtered test cases across $6$ programming languages. The benchmark emphasizes diversity, correctness, and challenge by retaining only cases that stump strong LLMs and undergo human annotation.
    \item Starting with 500k examples, a strict rejection sampling process ensured high quality by filtering repositories based on criteria such as having unit-test files, reference answer files, aligned code signatures and answers, compatible environment configurations, and standardized file names. The resulting high-quality dataset is used to fine-tune base LLMs to obtain \coder{}, enhancing their ability to understand and reason about multi-file codebases with clear dependencies and reproducible environments.
    \item Our systematic evaluation of over 40+ LLMs on \bench{} led to the creation of a dynamic leaderboard to track model performance, with extensive experiments demonstrating that \bench{} effectively measures the alignment between model-generated responses and human preferences.
\end{itemize}

\section{Automatic and Dynamic Pipeline for Repository-based Code Reflection}

\paragraph{Repository Code Data File Structure}Following Aider Code Edit Benchmark \citep{aiderCodeEditing} and Aider Polyglot Benchmark \citep{aiderpolyglotCodeEditing}, we collect 702 coding problems from  Exercism\footnote{\url{https://exercism.org/}} for C++, Go, Java, JavaScript, Python, and Rust, available at different repositories\footnote{\url{https://github.com/exercism}}. Then we remove the 225 coding problems in the Aider Polyglot Benchmark \cite{aiderpolyglotCodeEditing} tests and some erroneous data, and we retain about 473 coding problems to keep the file structure of newly generated code problems the same as these code problems. However, this code data may contain redundancy, as the purpose of these coding problems is to help people complete courses and become proficient in programming languages. Illustrated in \autoref{fig:file_structure_examples} in \autoref{appendix_polyglot_repository_code_file_structure}, we design a standardized repository code data file structure and streamline the file structure to the minimum size while ensuring normal evaluation compared to them. These data are used to guide the LLMs to generate the repository code file structure format of the newly generated coding problems but do not participate in guiding the content of the new coding problems illustrated in \autoref{fig:framework}.

\paragraph{Seed Code Data} To build \bench{} with high quality and diversity, we collect six programming languages (Python, Java, Go, Rust, C++, JavaScript) code from multiple public sources, like GitHub, Hugging Face, and Reddit.

\paragraph{Multiple Turn Dialogue Data Generation} We generate the program topics, definitions, unit tests, then reference answers sequencely and continue each step after the previous dialogue for consistency. To balance diversity and difficulty, a ``creative'' LLM randomly drawn from a mixed LLM stream produces topics and problem definitions, while multiple ``reasoning'' LLM, also randomly selected, generate unit tests and reference solutions.

\paragraph{Cross-execution Verification} We generate one program topic and definition but multiple unit-tests and multiple reference-answers for one coding program. unit-test and reference-answer pairs are sandbox cross-executed, and abnormal samples are dropped. For each program, we keep the unit test with the lowest pass rate and the reference answer with the highest; any 0\%–pass cases undergo manual inspection. Each curated coding program are then organized into a repository.

\section{\bench~Benchmark}

\paragraph{Selection of Executable Program}
Using our automated pipeline, we generate 100K coding program repository cases. We run them in a sandbox, including environment setup, compilation and testing, discarding any case that all LLMs can passe unit tests for too easy or exceeded 180s running time for too slow. Finally, we minimized same topic overlap per language to maximize diversity and retained 10K high-difficulty, high-correctness cases.

\paragraph{Selection of Difficult Problems}
For high correctness and difficulty of \bench{}, we generate the code signature based on the reference answer to organize coding program cases into a test repository following the file structure in \autoref{fig:file_structure_examples}. Then 10 selected mainstream strong reasoning LLMs will write the answer and each LLM has one chance to modify its answer if its initial answer not pass the unit test for some error. For each code program case, we collect the results of 10 LLMs and the result can be ``success'', ``failure-success'' and ``failure-failure''. 

If all 10 LLMs get ``success'', we think the code program case is easy and will discard it. If all 10 LLMs get ``failure-success'' or several model get ``failure-failure'', but more than half of the LLMs can still complete the task, we think this part of these code program cases has a certain degree of difficulty but cannot test all LLMs. Therefore, we will keep some data to test weaker LLMs. If more than half of the LLMs  only get ``failure-failure'' but there are still LLMs that can complete it, we think these code program cases have a high degree of difficulty and are more suitable for evaluating most LLMs. If there are some code program cases that all 10 LLMs get ``failure-failure'', we will keep them and mark them with special marks to focus on checking in the subsequent stages. The data that passes the inspection will be considered the most difficult part of the entire evaluation data. Finally, we discard nearly 8k data and only keep 2300 code program cases with high quality, high difficulty, high diversity.

\paragraph{Human Annotation}
To ensure the quality of \bench{}, we employ 8 graduate students and provide them with a complete code running sandbox environment and ensure the integrity of the data. Each person will complete the annotation of nearly 300 code program cases with the assistance of LLMs. The inspection tasks for each code program case are to check the rationality of the code program case, check the code environment configuration file, check the code file structure, check the reference answer and unit test. In the end, we retained 1,888 test code program cases as the final version of the test data.

\paragraph{Dataset Statistics}

%%%%%%%%%%%%%%%%%%%%%%%%%%%%%%%%%%%%%%%%%%%%%%%%%%%%%%%%%%%%%%%%%%%%%%
\begin{figure}[t!]
  \centering
  \includegraphics[width=1.0\columnwidth]{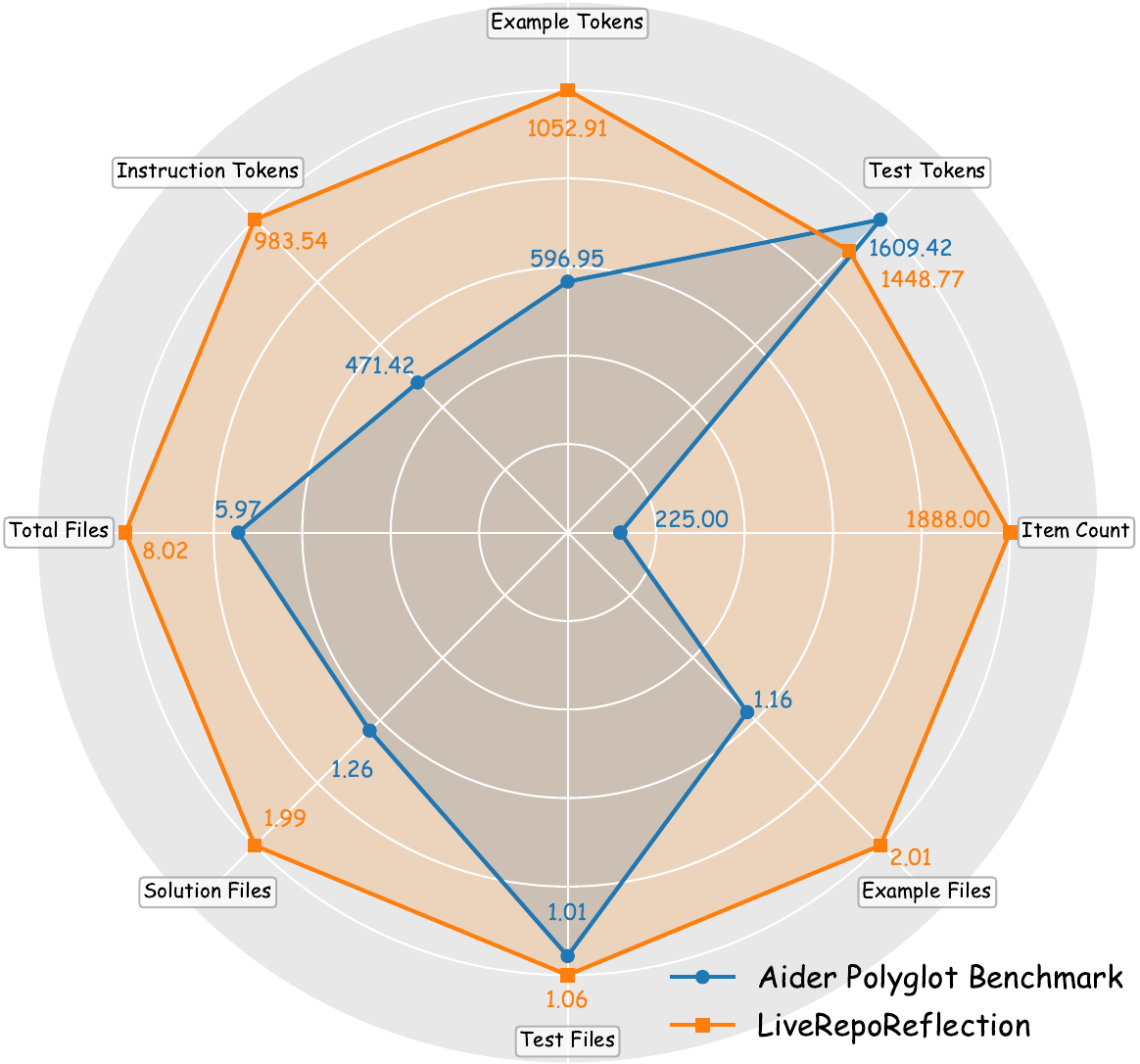}
  \caption{Comparison of dataset scales and structures between Aider Polyglot Benchmark (blue) and \bench{} (orange).  We display eight metrics: (1) problem count; (2) average tokens in test suites; (3) average example‐context tokens; (4) average instruction tokens; and (5–8) average files per repository (total, solution, test, example). These metrics highlight that \bench{} substantially exceeds Aider Polyglot Benchmark in problem coverage, contextual richness, and real-world, multi-file layout complexity.}
  \label{fig:radar}
  \vspace{-15pt}
\end{figure}
%%%%%%%%%%%%%%%%%%%%%%%%%%%%%%%%%%%%%%%%%%%%%%%%%%%%%%%%%%%%%%%%%%%%%%

Figure~\ref{fig:radar} presents a head-to-head comparison of \bench{} (orange) against the Aider Polyglot Benchmark (blue) over eight key dimensions.  First, \bench{} comprises 1,888 problems—more than 8× the 225 in Aider Polyglot Benchmark, enabling far broader coverage.  In terms of test‐suite length, our benchmarks average 1,448.8 tokens versus 1,609.4 in Aider Polyglot Benchmark, reflecting more concise, targeted checks.  Example contexts in \bench{} average 1,052.9 tokens compared to 596.9 (≈1.8×), and instructional contexts average 983.5 tokens versus 471.4 (≈2.1×), underscoring richer problem descriptions. Structurally, each \bench{} repository contains an average of 8.02 files (vs.\ 5.97), including 2.00 solution files (vs.\ 1.26), 1.06 test files (vs.\ 1.01), and 2.01 example files (vs.\ 1.16).  This expand, multi-file layout more faithfully mirrors real-world codebases and challenges models to navigate complex project structures.  Overall, \bench{} delivers substantially greater scale, depth, and structural complexity, making it a more rigorous and realistic testbed for modern code-generation systems. And we provide language distribution comparison in \autoref{tab:language_distribution} in \autoref{language_distribution_comparison}.

%%%%%%%%%%%%%%%%%%%%%%%%%%%%%%%%%%%%%%%%%%%%%%%%%%%%%%%%%%%%%%%%%%%%%%%%%%%%%%%%%%%%%
\begin{table*}[t!]
  \centering
  \small
  \renewcommand{\arraystretch}{1.1}
  \resizebox{\textwidth}{!}{%
    \begin{tabular}{ll|cccc|cccc|cccc|cccc|cccc|cccc|cccc|}
      \toprule
       \multirow{2}{*}{Model} & \multirow{2}{*}{Size} & \multicolumn{4}{c}{Python} & \multicolumn{4}{c}{Java} & \multicolumn{4}{c}{Cpp} & \multicolumn{4}{c}{Rust} & \multicolumn{4}{c}{Go} & \multicolumn{4}{c}{Javascript} & \multicolumn{4}{c}{All} \\
        &  & P1 & P2 & WF & FW & P1 & P2 & WF & FW & P1 & P2 & WF & FW & P1 & P2 & WF & FW & P1 & P2 & WF & FW & P1 & P2 & WF & FW & P1 & P2 & WF & FW \\
      \midrule
      \rowcolor{HeaderColor}\multicolumn{30}{c}{\textit{Closed-Source LLMs}} \\
      \midrule
      \rowcolor{ClosedColor} \small Claude-3-5-Haiku-20241022 & \faLock{} & 11.5 & 33.7 & 97.8 & 65.9 & 12.8 & 22.4 & 96.4 & 42.9 & 4.2 & 13.3 & 89.5 & 68.4 & 1.4 & 5.7 & \bf 100.0 & 75.0 & 2.7 & 16.9 & 99.5 & 83.8 & 1.7 & 50.0 & 95.0 & 96.6 & 7.8 & 24.4 & 97.5 & 68.1 \\
      \rowcolor{ClosedColor} \small Claude-3-5-Sonnet-20240620 & \faLock{} & 12.1 & 34.6 & 98.3 & 65.1 & 12.0 & 23.6 & 97.6 & 49.2 & 3.5 & 8.4 & 84.6 & 58.3 & 1.4 & 6.6 & 99.5 & 78.6 & 3.7 & 16.6 & 98.3 & 77.6 & 1.7 & 45.0 & 98.3 & 96.2 & 8.1 & 24.5 & 97.3 & 67.0 \\
      \rowcolor{ClosedColor} \small Claude-3-5-Sonnet-20241022 & \faLock{} & 14.4 & 45.7 & 98.2 & 68.5 & 10.8 & 30.4 & 99.2 & 64.5 & 4.2 & 14.7 & 99.3 & 71.4 & 2.8 & 7.5 & 99.5 & 62.5 & 5.7 & 23.6 & 95.5 & 75.8 & 1.7 & 53.3 & 78.3 & 96.8 & 9.6 & 32.6 & 97.4 & 70.6 \\
      \rowcolor{ClosedColor} \small Claude-3-7-Sonnet-20250219 & \faLock{} & 17.6 & 51.1 & 99.9 & 65.6 & \bf 14.0 & \bf 33.2 & \bf 100.0 & 57.8 & 6.3 & 23.1 & 98.6 & 72.7 & 3.3 & 6.6 & \bf 100.0 & 50.0 & 6.5 & 25.8 & \bf 100.0 & 75.0 & 1.7 & 80.0 & 98.3 & 97.9 & 11.8 & 37.1 & 99.8 & 68.3 \\
      \rowcolor{ClosedColor} \small Claude-3-7-Sonnet-20250219-Thinking & \faLock{} & 14.1 & 40.7 & 99.9 & 65.3 & 12.8 & 25.6 & 99.2 & 50.0 & 4.9 & 20.3 & 99.3 & 75.9 & 3.3 & 8.0 & \bf 100.0 & 58.8 & 6.0 & 18.4 & 99.8 & 67.6 & 1.7 & 43.3 & \bf 100.0 & 96.0 & 9.9 & 28.8 & 99.7 & 65.6 \\
      \rowcolor{ClosedColor} \small Doubao-1-5-Thinking-pro-m-250415 & \faLock{} & 10.1 & 33.5 & 99.0 & 69.8 & 2.4 & 16.4 & 99.6 & 85.4 & 4.2 & 14.7 & 97.2 & 71.4 & 0.5 & 3.3 & \bf 100.0 & 85.7 & 1.5 & 6.7 & 99.5 & 77.8 & 16.7 & 75.0 & \bf 100.0 & 77.7 & 5.9 & 22.0 & 99.2 & 73.1 \\
      \rowcolor{ClosedColor} \small Gemini-2.0-Flash & \faLock{} & 5.6 & 26.6 & 99.0 & 78.9 & 5.6 & 16.0 & 96.8 & 65.0 & 2.8 & 8.4 & 97.2 & 66.7 & 1.9 & 3.3 & \bf 100.0 & 42.9 & 0.5 & 2.7 & 98.8 & 81.8 & 0.0 & 23.3 & 96.7 & \bf 100 & 3.7 & 16.0 & 98.6 & 76.8 \\
      \rowcolor{ClosedColor} \small Gemini-2.5-Flash-preview-04-17 & \faLock{} & 1.7 & 19.1 & 59.5 & 91.1 & 1.6 & 10.8 & 82.8 & 85.2 & 2.8 & 16.1 & 57.3 & 82.6 & 2.4 & 5.2 & 97.6 & 54.5 & 2.7 & 8.2 & 80.1 & 66.7 & 0.0 & 10.0 & 76.7 & \bf 100 & 2.0 & 13.6 & 71.7 & 85.2 \\
      \rowcolor{ClosedColor} \small Gemini-2.5-pro-preview-05-06 & \faLock{} & 0.5 & 13.5 & \bf 100.0 & \bf 96.4 & 0.8 & 6.0 & \bf 100.0 & 86.7 & 0.7 & 4.9 & \bf 100.0 & \bf 85.7 & 0.0 & 3.8 & \bf 100.0 & 100.0 & 1.5 & 3.2 & 99.5 & 53.8 & 0.0 & 5.0 & \bf 100.0 & \bf 100 & 0.7 & 8.3 & 99.9 & \bf 91.7 \\
      \rowcolor{ClosedColor} \small GPT-4o-mini-2024-07-18 & \faLock{} & 7.6 & 22.9 & 98.3 & 67.0 & 2.8 & 14.0 & 95.6 & 80.0 & 2.8 & 7.7 & 85.3 & 63.6 & 1.9 & 2.8 & 98.6 & 33.3 & 0.2 & 4.7 & 96.8 & \bf 94.7 & 41.7 & 75.0 & \bf 100.0 & 44.4 & 5.5 & 16.1 & 96.7 & 66.1 \\
      \rowcolor{ClosedColor} \small GPT-4o-2024-11-20 & \faLock{} & 9.1 & 30.4 & 99.6 & 69.9 & 8.8 & 21.6 & 99.6 & 59.3 & 3.5 & 14.7 & \bf 100.0 & 76.2 & 2.8 & 6.6 & \bf 100.0 & 57.1 & 2.0 & 11.2 & \bf 100.0 & 82.2 & 46.7 & 68.3 & \bf 100.0 & 31.6 & 7.6 & 22.5 & 99.8 & 66.0 \\
      \rowcolor{ClosedColor} \small GPT-4.1-2025-04-14 & \faLock{} & 10.6 & 45.7 & \bf 100.0 & 76.8 & 3.6 & 27.6 & \bf 100.0 & 87.0 & 4.9 & 26.6 & \bf 100.0 & 81.6 & 3.3 & 9.4 & \bf 100.0 & 65.0 & 2.2 & 14.9 & 99.8 & 85.0 & 28.3 & 36.7 & \bf 100.0 & 22.9 & 7.2 & 30.9 & 99.9 & 76.7 \\
      \rowcolor{ClosedColor} \small GPT-4.1-mini-2025-04-14 & \faLock{} & 10.9 & 38.8 & 99.9 & 72.0 & 4.0 & 17.6 & \bf 100.0 & 77.3 & 7.7 & 21.7 & \bf 100.0 & 64.5 & 4.2 & 7.1 & \bf 100.0 & 40.0 & 4.7 & 19.1 & 99.8 & 75.3 & 50.0 & \bf 85.0 & \bf 100.0 & 41.2 & 8.9 & 28.4 & 99.9 & 68.7 \\
      \rowcolor{ClosedColor} \small GPT-4.1-nano-2025-04-14 & \faLock{} & 3.2 & 12.8 & 75.7 & 75.2 & 0.8 & 5.2 & 47.2 & 84.6 & 2.1 & 11.2 & 53.8 & 81.2 & 0.5 & 0.9 & 92.9 & 50.0 & 0.2 & 4.0 & 73.4 & 93.8 & 10.0 & 56.7 & 65.0 & 0.8 & 2.1 & 9.9 & 71.4 & 79.0 \\
      \rowcolor{ClosedColor} \small GPT-4.5-preview-2025-02-27 & \faLock{} & 11.1 & 40.1 & \bf 100.0 & 72.3 & 3.2 & 23.6 & \bf 100.0 & 86.4 & 7.0 & 19.6 & \bf 100.0 & 64.3 & 3.8 & 8.5 & \bf 100.0 & 55.6 & 3.2 & 14.4 & \bf 100.0 & 77.6 & 21.7 & 33.3 & \bf 100.0 & 34.8 & 7.6 & 27.1 & \bf 100.0 & 72.1 \\
      \rowcolor{ClosedColor} \small Grok-3 & \faLock{} & 12.2 & 40.6 & \bf 100.0 & 70.0 & 4.8 & 22.8 & \bf 100.0 & 78.9 & 5.6 & 15.4 & 99.3 & 63.6 & 3.3 & 8.0 & \bf 100.0 & 58.8 & 4.5 & 20.1 & \bf 100.0 & 77.8 & 0.0 & 60.0 & \bf 100.0 & \bf 100 & 7.7 & 28.9 & 99.9 & 73.4 \\
      \rowcolor{ClosedColor} \small Grok-3-fast & \faLock{} & 14.3 & 42.4 & 99.9 & 66.4 & 4.8 & 25.2 & \bf 100.0 & 81.0 & 5.6 & 16.1 & 99.3 & 65.2 & 4.7 & 9.4 & \bf 100.0 & 50.0 & 5.7 & 20.1 & 99.8 & 71.6 & 0.0 & 50.0 & \bf 100.0 & \bf 100 & 9.0 & 29.9 & 99.8 & 69.9 \\
      \rowcolor{ClosedColor} \small Grok-3-mini & \faLock{} & 9.9 & 38.0 & 96.7 & 74.0 & 3.6 & 22.4 & 99.6 & 83.9 & 9.1 & 21.7 & 98.6 & 58.1 & 3.3 & 8.0 & \bf 100.0 & 58.8 & 1.5 & 11.9 & 99.5 & 87.5 & 1.7 & 28.3 & \bf 100.0 & 94.0 & 6.2 & 25.5 & 98.3 & 75.7 \\
      \rowcolor{ClosedColor} \small Grok-3-mini-fast & \faLock{} & 8.9 & 37.2 & 96.3 & 76.1 & 1.2 & 24.0 & \bf 100.0 & 95.0 & 5.6 & 19.6 & \bf 100.0 & 71.4 & 1.9 & 7.5 & \bf 100.0 & 75.0 & 3.2 & 14.9 & \bf 100.0 & 78.3 & 0.0 & 23.3 & \bf 100.0 & \bf 100 & 5.3 & 25.6 & 98.4 & 79.1 \\
      \rowcolor{ClosedColor} \small o1-mini-2024-09-12 & \faLock{} & 12.2 & 41.2 & 98.4 & 70.4 & 2.8 & 18.0 & 92.0 & 84.4 & 5.6 & 23.1 & 82.5 & 75.8 & 2.4 & 4.7 & 97.6 & 50.0 & 4.2 & 16.9 & 96.5 & 75.0 & \bf 55.0 & \bf 85.0 & 95.0 & 35.3 & 9.0 & 28.9 & 95.8 & 68.8 \\
      \rowcolor{ClosedColor} \small o3-mini-2025-01-31 & \faLock{} & 18.8 & 50.9 & \bf 100.0 & 63.1 & 4.8 & 31.6 & 98.8 & 84.8 & \bf 10.5 & 32.9 & \bf 100.0 & 68.1 & 4.7 & 8.5 & \bf 100.0 & 44.4 & 4.2 & 18.9 & \bf 100.0 & 77.6 & 26.7 & 75.0 & \bf 100.0 & 64.4 & 11.9 & 36.1 & 99.8 & 67.2 \\
      \rowcolor{ClosedColor} \small o4-mini-2025-04-16 & \faLock{} & \bf 20.4 & \bf 54.9 & 99.8 & 62.9 & 4.4 & \bf 33.2 & \bf 100.0 & 86.7 & \bf 10.5 & \bf 41.3 & 93.0 & 74.6 & \bf 6.6 & \bf 10.8 & \bf 100.0 & 39.1 & \bf 7.7 & \bf 29.8 & 99.3 & 74.2 & 45.0 & 50.0 & 98.3 & 10.0 & \bf 14.0 & \bf 40.5 & 99.2 & 65.4 \\
      \midrule
      \rowcolor{HeaderColor}\multicolumn{30}{c}{\textit{Open-Source LLMs}} \\
      \midrule
      \rowcolor{OpenColor} \small Qwen3-0.6B-Instruct Think & 0.6B & 0.6 & 1.1 & 83.2 & 44.4 & 0.8 & 0.8 & 72.4 & 0.0 & 0.0 & 0.0 & 66.4 & 0.0 & 0.0 & 0.0 & 94.8 & 0.0 & 0.0 & 0.0 & 68.0 & 0.0 & 3.3 & 5.0 & 68.3 & 34.0 & 0.5 & 0.7 & 78.1 & 35.7 \\
      \rowcolor{OpenColor} \small Qwen3-0.6B-Instruct Chat & 0.6B & 0.4 & 0.5 & 90.7 & 25.0 & 0.4 & 0.4 & 72.4 & 0.0 & 0.0 & 0.0 & 93.7 & 0.0 & 0.0 & 0.0 & 97.2 & 0.0 & 0.0 & 0.0 & 93.8 & 0.0 & 0.0 & 0.0 & 58.3 & 0.0 & 0.2 & 0.3 & 88.9 & 20.0 \\
      \rowcolor{OpenColor} \small Qwen3-1.7B-Instruct Chat & 1.7B & 1.2 & 2.1 & 94.6 & 41.2 & 1.6 & 1.6 & 92.0 & 0.0 & 0.0 & 0.0 & \bf 100.0 & 0.0 & 0.0 & 0.0 & 98.6 & 0.0 & 0.0 & 0.5 & 99.0 & 100.0 & 8.3 & 21.7 & 91.7 & 61.8 & 1.0 & 1.9 & 96.0 & 47.2 \\
      \rowcolor{OpenColor} \small Qwen3-1.7B-Instruct Think & 1.7B & 2.8 & 6.8 & 79.6 & 58.9 & 1.2 & 1.2 & 50.4 & 0.0 & 0.0 & 0.0 & 61.5 & 0.0 & 0.0 & 0.0 & 78.8 & 0.0 & 0.5 & 0.5 & 62.3 & 0.0 & 0.0 & 6.7 & 38.3 & \bf 100 & 1.5 & 3.4 & 69.3 & 56.9 \\
      \rowcolor{OpenColor} \small Qwen3-4B-Instruct Chat & 4B & 4.0 & 10.9 & 98.4 & 62.9 & 1.2 & 2.4 & \bf 100.0 & 50.0 & 2.1 & 4.9 & \bf 100.0 & 57.1 & 0.0 & 0.5 & \bf 100.0 & \bf 100.0 & 0.7 & 2.5 & \bf 100.0 & 70.0 & 5.0 & 21.7 & 100.0 & 90.8 & 2.4 & 6.7 & 99.3 & 64.3 \\
      \rowcolor{OpenColor} \small Qwen3-4B-Instruct Think & 4B & 3.7 & 17.8 & 92.2 & 79.5 & 0.8 & 4.0 & 72.4 & 80.0 & 2.1 & 7.0 & 67.8 & 70.0 & 0.0 & 0.5 & 95.3 & \bf 100.0 & 1.5 & 5.0 & 86.6 & 70.0 & 0.0 & 21.7 & 78.3 & \bf 100 & 2.2 & 10.6 & 86.4 & 79.5 \\
      \rowcolor{OpenColor} \small Qwen3-8B-Instruct Think & 8B & 4.4 & 22.9 & 92.3 & 80.9 & 3.6 & 8.8 & 80.0 & 59.1 & 1.4 & 10.5 & 73.4 & \bf 86.7 & 0.0 & 1.9 & 98.1 & \bf 100.0 & 1.7 & 7.9 & 84.4 & 78.1 & 5.0 & 16.7 & 81.7 & 77.0 & 3.0 & 14.4 & 87.9 & 79.0 \\
      \rowcolor{OpenColor} \small Qwen3-8B-Instruct Chat & 8B & 4.3 & 15.6 & 99.9 & 72.7 & 9.6 & 19.2 & 99.6 & 50.0 & 1.4 & 4.2 & \bf 100.0 & 66.7 & 0.0 & 0.5 & \bf 100.0 & \bf 100.0 & 0.5 & 4.0 & 99.3 & \bf 87.5 & 20.0 & 43.3 & 98.3 & 53.8 & 4.0 & 11.9 & 99.7 & 66.7 \\
      \rowcolor{OpenColor} \small OpenCoder-8B-Instruct & 8B & 0.2 & 0.2 & \bf 100.0 & 0.0 & 0.8 & 0.8 & \bf 100.0 & 0.0 & 0.0 & 0.0 & \bf 100.0 & 0.0 & 0.0 & 0.0 & \bf 100.0 & 0.0 & 0.0 & 0.0 & \bf 100.0 & 0.0 & 0.0 & 0.0 & 100.0 & 0.0 & 0.2 & 0.2 & \bf 100.0 & 0.0 \\
      \rowcolor{OpenColor} \small Seed-Coder-8B-Instruct & 8B & 4.4 & 13.3 & 91.6 & 67.0 & 2.0 & 11.6 & 47.2 & 82.8 & 2.1 & 3.5 & 55.9 & 40.0 & 2.4 & 3.8 & 98.1 & 37.5 & 1.0 & 3.2 & 82.1 & 69.2 & 1.7 & 13.3 & 93.3 & 87.2 & 2.9 & 9.1 & 81.8 & 68.6 \\
      \rowcolor{OpenColor} \small Qwen3-14B-Instruct Chat & 14B & 4.8 & 20.0 & 99.4 & 76.2 & 4.4 & 15.6 & 99.2 & 71.8 & 3.5 & 6.3 & 98.6 & 44.4 & 0.5 & 2.8 & 99.5 & 83.3 & 1.2 & 8.4 & 99.0 & 85.3 & 1.7 & 23.3 & 96.7 & 92.7 & 3.3 & 14.1 & 99.2 & 76.7 \\
      \rowcolor{OpenColor} \small Qwen3-14B-Instruct Think & 14B & 4.8 & 26.0 & 87.7 & 81.7 & 2.0 & 16.4 & 71.6 & 87.8 & 1.4 & 10.5 & 69.2 & \bf 86.7 & 1.4 & 3.3 & 92.9 & 57.1 & 3.5 & 9.2 & 85.6 & 62.2 & 1.7 & 11.7 & 71.7 & 85.5 & 3.4 & 16.9 & 83.8 & 80.0 \\
      \rowcolor{OpenColor} \small Qwen3-30B-A3B-Instruct Think & 3/30B & 6.3 & 29.3 & 91.5 & 78.3 & 1.6 & 15.2 & 76.8 & 89.5 & 2.8 & 11.9 & 62.2 & 76.5 & 2.4 & 4.2 & 95.3 & 44.4 & 2.7 & 12.9 & 84.1 & 78.8 & 13.3 & 35.0 & 73.3 & 62.0 & 4.4 & 20.0 & 85.6 & 77.7 \\
      \rowcolor{OpenColor} \small Qwen3-30B-A3B-Instruct Chat & 3/30B & 7.7 & 23.0 & 98.8 & 66.7 & 0.8 & 13.6 & \bf 100.0 & \bf 94.1 & 3.5 & 7.7 & 99.3 & 54.5 & 0.9 & 0.9 & 99.5 & 0.0 & 1.7 & 6.9 & 1\bf 00.0 & 75.0 & 8.3 & 40.0 & 100.0 & 79.3 & 4.4 & 15.3 & 99.4 & 70.8 \\
      \rowcolor{OpenColor} \small Qwen2.5-Coder-32B-Instruct & 32B & 6.6 & 22.7 & 98.4 & 71.0 & 3.2 & 10.8 & 93.6 & 70.4 & 2.1 & 6.3 & 98.6 & 66.7 & 0.9 & 4.7 & 98.6 & 80.0 & 1.2 & 6.5 & 98.5 & 80.8 & 1.7 & 38.3 & 91.7 & 95.6 & 3.9 & 14.9 & 97.6 & 74.0 \\
      \rowcolor{OpenColor} \small Qwen3-32B-Instruct Think & 32B & 5.7 & 32.0 & 86.1 & 82.1 & 6.8 & 17.6 & 70.4 & 61.4 & 3.5 & 18.2 & 51.7 & 80.8 & 1.4 & 4.7 & 97.2 & 70.0 & 2.7 & 10.7 & 78.7 & 74.4 & 15.0 & 38.3 & 63.3 & 69.4 & 4.9 & 21.6 & 80.3 & 77.5 \\
      \rowcolor{OpenColor} \small Qwen3-32B-Instruct Chat & 32B & 7.7 & 27.9 & 95.1 & 72.5 & 6.4 & 19.6 & 89.2 & 67.3 & 5.6 & 14.7 & 83.2 & 61.9 & 1.4 & 2.4 & 98.6 & 40.0 & 1.7 & 8.9 & 93.3 & 80.6 & \bf 28.3 & 60.0 & 85.0 & 52.8 & 6.0 & 19.9 & 93.1 & 69.7 \\
      \rowcolor{OpenColor} \small QwQ-32B & 32B & 2.1 & 21.8 & 92.6 & \bf 90.5 & 1.6 & 8.4 & 94.4 & 81.0 & 3.5 & 13.3 & 91.6 & 73.7 & 1.4 & 3.3 & 98.1 & 57.1 & 1.0 & 5.2 & 94.8 & 81.0 & 0.0 & 8.3 & 76.7 & \bf 100 & 1.7 & 13.3 & 93.3 & \bf 86.9 \\
      \rowcolor{OpenColor} \small DeepSeek-V3 & 37/671B & 11.1 & 40.1 & 99.3 & 72.3 & 11.2 & 29.6 & 99.6 & 62.2 & 6.3 & 15.4 & \bf 100.0 & 59.1 & 3.3 & 7.1 & \bf 100.0 & 53.3 & 2.7 & 18.4 & 98.5 & 85.1 & 0.0 & 18.3 & 96.7 & \bf 100 & 7.7 & 27.8 & 99.2 & 72.2 \\
      \rowcolor{OpenColor} \small DeepSeek-R1 & 37/671B & 9.6 & 37.6 & 99.9 & 74.4 & 6.4 & 26.0 & 99.6 & 75.4 & 7.0 & \bf 22.4 & 98.6 & 68.8 & \bf 3.8 & \bf 9.4 & \bf 100.0 & 60.0 & \bf 4.5 & 17.6 & 99.3 & 74.6 & 0.0 & 46.7 & 100.0 & \bf 100 & 6.9 & 27.8 & 99.6 & 75.0 \\
      \rowcolor{OpenColor} \small DeepSeek-V3-250324 & 37/671B & 12.7 & 39.9 & 99.8 & 68.2 & \bf 16.0 & \bf 30.8 & \bf 100.0 & 48.1 & 4.9 & 12.6 & \bf 100.0 & 61.1 & 2.4 & 7.5 & \bf 100.0 & 68.8 & 3.7 & \bf 19.1 & 99.5 & 80.5 & 0.0 & 15.0 & 100.0 & \bf 100 & \bf 9.1 & 27.8 & 99.8 & 67.4 \\
      \rowcolor{OpenColor} \small DeepSeek-R1-Distill-Llama-8B & 8B & 1.0 & 2.8 & 76.2 & 65.2 & 0.4 & 0.8 & 38.0 & 50.0 & 0.0 & 2.8 & 42.7 & 100.0 & 0.0 & 0.0 & 85.8 & 0.0 & 0.0 & 0.2 & 53.3 & 100.0 & 5.0 & 6.7 & 25.0 & 25.0 & 0.6 & 1.8 & 63.2 & 64.7 \\
      \rowcolor{OpenColor} \small DeepSeek-R1-Distill-Qwen-14B & 14B & 3.9 & 13.0 & 94.0 & 70.1 & 0.4 & 1.6 & 81.6 & 75.0 & 0.7 & 2.8 & 94.4 & 75.0 & 0.0 & 0.0 & 97.2 & 0.0 & 0.5 & 1.0 & 81.4 & 50.0 & 0.0 & 5.0 & 48.3 & 100.0 & 1.9 & 6.5 & 88.6 & 70.5 \\
      \rowcolor{OpenColor} \small Llama3.1-8B-Instruct & 8B & 1.3 & 4.4 & 87.2 & 69.4 & 1.2 & 7.6 & 72.0 & 84.2 & 0.7 & 0.7 & 63.6 & 0.0 & 0.0 & 0.0 & 92.9 & 0.0 & 0.5 & 0.5 & 83.6 & 0.0 & 6.7 & 15.0 & 70.0 & 55.6 & 1.1 & 3.5 & 82.7 & 68.7 \\
      \midrule
      \rowcolor{OpenColor} \small \coder & 32B & \bf 13.5 & \bf 40.4 & 99.8 & 66.5 & 8.4 & 22.0 & 99.6 & 61.8 & 5.6 & 15.4 & \bf 100.0 & 63.6 & 1.4 & 3.3 & \bf 100.0 & 57.1 & \bf 4.5 & 17.4 & \bf 100.0 & 74.3 & 15.0 & \bf 78.3 & \bf 100.0 & 80.8 & 9.0 & \bf 28.2 & 99.8 & 68.0 \\
      \bottomrule
    \end{tabular}%
  }%
  \caption{Main Results for `\new`.}
  \label{tab:results_whole}
  \vspace{-20pt}
\end{table*}
%%%%%%%%%%%%%%%%%%%%%%%%%%%%%%%%%%%%%%%%%%%%%%%%%%%%%%%%%%%%%%%%%%%%%%%%%%%%%%%%%%%%%

\begin{figure}[t]
\begin{center}
    \includegraphics[width=0.8\columnwidth]{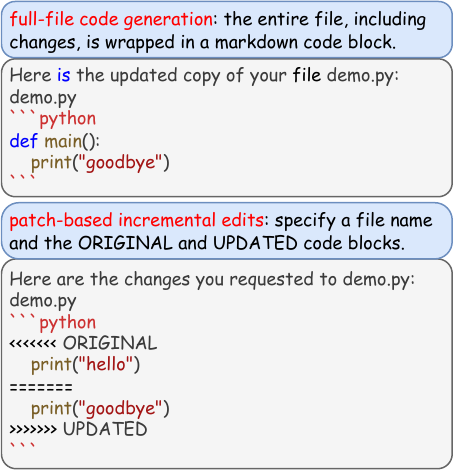}
    \caption{Two evaluation edit format of \bench{}.}
    \label{fig:edit_format}
    \vspace{-20pt}
\end{center}
\end{figure}

\section{Training of \coder{}}

\paragraph{Source Data Creation and Quality-Based Reject Sampling}
We collect approximately 500,000 code examples (including 400,000 newly automatically and dynamically generated ones) to construct the \instruction{} corpus. To ensure data quality, strict rejection sampling was applied to all examples, and only repositories meeting all five of the following criteria were retained as a high-quality fine-tuning dataset:  1) At least one unit test file;  2) At least one reference answer file;  3) The number of code signature files matches the number of reference answer files;  4) Environment configuration files correspond with the declared programming language;  5) File naming and extensions are standardized and free of anomalies.

\paragraph{Quality Scoring Mechanism}
Following reject sampling, we employ LLM as a judge and the following sophisticated scoring checklist to quantitatively assess each code program. The composite score \(S(p)\) for a code program \(p\) is formulated as a weighted linear combination of five key metrics:  $S(p)=\sum_{i\in\{\mathrm{1,2,3,4,5}\}}w_i\,S_i$ , where $w=(0.3,0.2,0.2,0.15,0.15)$. The five key points are listed as:
(1) $S_{1}=S_{exec}$: Executability score derived from unit-test pass rate and the absence of compilation or interpretation errors, reflecting functional correctness. (2) $S_{2}=S_{nov}$: Novelty score measuring code similarity against existing repositories and semantic diversity in problem descriptions. (3) $S_3=S_{dif}$: Difficulty score inferred from low test pass rates and the extent of boundary and edge-case coverage, to select challenging tasks. (4) $S_4=S_{style}$: Code style score based on static analysis tools (e.g., Black, ESLint) that evaluate adherence to naming conventions, code comments, and formatting. (5) $S_5=S_{ppl}$: Inverse perplexity on reference answers, indicating code that is more challenging for the model to predict and thus potentially more informative.  

\paragraph{\bench{}~Decontamination}
We split the top 10,000 code problems ranked by the scoring function into character 5-gram fragments, use the MinHash algorithm to generate compact signatures, and then utilize an LSH index to efficiently filter candidate texts exhibiting Jaccard similarity greater than 0.8 with the test set. Finally, we perform exact matching verification to ensure the effective and accurate removal of duplicate or highly similar data, thereby maintaining the purity and high quality of the dataset. A total of 8,702 high-quality training code program cases, strictly decontaminated with the original 1,888 test sets in \bench{}.

\paragraph{Multi-turn Interaction for \instruction{} Generation}

To achieve diverse multi-turn interaction styles, we simulated 840,839 multi-turn coding dialogues using four top models (claude-3-7-sonnet-20250219, qwen3-235b-a22b-instruct, o1-mini, o4-mini), each run four times in two formats (full-file generation and patch-based edits). Training examples are filtered by task importance as follows: 1) Direct generation (40\%): produce a complete solution from a prompt. 2) Error-driven repair (40\%): iteratively fix code based on compiler/runtime errors. 3) Style standardization (10\%): refactor code to meet linting/style guidelines. 4) Dialogue summarization (10\%): condense multi-turn interactions into concise overviews.

\begin{table*}[t]
  \centering
  \small
  \renewcommand{\arraystretch}{1.1}
  \resizebox{\textwidth}{!}{%
    \begin{tabular}{ll|cccc|cccc|cccc|cccc|cccc|cccc|cccc|}
      \toprule
       \multirow{2}{*}{Model} & \multirow{2}{*}{Size} & \multicolumn{4}{c}{Python} & \multicolumn{4}{c}{Java} & \multicolumn{4}{c}{Cpp} & \multicolumn{4}{c}{Rust} & \multicolumn{4}{c}{Go} & \multicolumn{4}{c}{Javascript} & \multicolumn{4}{c}{All} \\
        &  & P1 & P2 & WF & FW & P1 & P2 & WF & FW & P1 & P2 & WF & FW & P1 & P2 & WF & FW & P1 & P2 & WF & FW & P1 & P2 & WF & FW & P1 & P2 & WF & FW \\
      \midrule
      \rowcolor{HeaderColor}\multicolumn{30}{c}{\textit{Closed-Source LLMs}} \\
      \midrule
      \rowcolor{ClosedColor} \small Claude-3-5-Haiku-20241022 & \faLock{} & 12.0 & 36.6 & 98.9 & 67.3 & 12.4 & 25.2 & 95.6 & 50.8 & 2.1 & 9.9 & 92.3 & 78.6 & 2.4 & 5.2 & 99.5 & 54.5 & 3.2 & 16.4 & 99.5 & 80.3 & 1.7 & 41.7 & \bf 100.0 & 95.9 & 8.0 & 25.4 & 98.2 & 68.5 \\
      \rowcolor{ClosedColor} \small Claude-3-5-Sonnet-20240620 & \faLock{} & 11.2 & 35.4 & 98.8 & 68.3 & 11.6 & 22.4 & 94.8 & 48.2 & 2.8 & 10.5 & 95.1 & 73.3 & 1.4 & 5.7 & 98.6 & 75.0 & 4.0 & 16.1 & 98.3 & 75.4 & 3.3 & 45.0 & \bf 100.0 & 92.7 & 7.7 & 24.6 & 97.9 & 68.6 \\
      \rowcolor{ClosedColor} \small Claude-3-5-Sonnet-20241022 & \faLock{} & 13.8 & 41.5 & \bf 99.4 & 66.8 & 14.4 & 28.8 & \bf 99.6 & 50.0 & 2.8 & 14.0 & 89.5 & 80.0 & 3.3 & 5.2 & \bf 100.0 & 36.4 & \bf 6.7 & 21.6 & \bf 99.8 & 69.0 & 1.7 & 31.7 & \bf 100.0 & 94.6 & 10.0 & 29.1 & \bf 98.8 & 65.8 \\
      \rowcolor{ClosedColor} \small Claude-3-7-Sonnet-20250219 & \faLock{} & 17.4 & 51.2 & 97.9 & 66.0 & \bf 14.8 & 29.2 & 95.6 & 49.3 & 5.6 & 22.4 & 88.1 & 75.0 & 3.8 & 6.6 & 98.1 & 42.9 & 5.7 & 21.3 & 98.0 & 73.3 & 0.0 & 70.0 & 93.3 & \bf 100.0 & 11.6 & 35.3 & 96.8 & 67.2 \\
      \rowcolor{ClosedColor} \small Claude-3-7-Sonnet-20250219-Thinking & \faLock{} & 14.1 & 45.2 & 98.9 & 68.7 & 13.2 & 29.2 & \bf 99.6 & 54.8 & 7.0 & 18.2 & 95.1 & 61.5 & 2.4 & 8.0 & 99.1 & 70.6 & 4.7 & 18.9 & 99.0 & 75.0 & 1.7 & 56.7 & 98.3 & 97.0 & 9.7 & 31.6 & 98.7 & 69.2 \\
      \rowcolor{ClosedColor} \small Doubao-1-5-Thinking-pro-m-250415 & \faLock{} & 9.9 & 30.9 & 93.4 & 68.0 & 2.0 & 14.4 & 91.2 & 86.1 & 5.3 & 8.4 & 58.0 & 36.4 & 0.9 & 1.9 & 75.9 & 50.0 & 0.7 & 7.2 & 93.8 & 89.7 & 0.0 & 0.0 & 90.0 & 0.0 & 5.2 & 17.8 & 88.6 & 70.6 \\
      \rowcolor{ClosedColor} \small Gemini-2.0-Flash & \faLock{} & 6.3 & 23.7 & 76.7 & 73.2 & 4.8 & 8.0 & 80.0 & 40.0 & 2.1 & 6.3 & 74.8 & 66.7 & 2.4 & 3.8 & 88.2 & 37.5 & 0.2 & 3.2 & 78.2 & 92.3 & 0.0 & 10.0 & 81.7 & \bf 100.0 & 3.9 & 13.2 & 78.8 & 70.8 \\
      \rowcolor{ClosedColor} \small Gemini-2.5-Flash-preview-04-17 & \faLock{} & 4.6 & 22.6 & 75.1 & 79.5 & 3.2 & 10.8 & 72.8 & 70.4 & 2.8 & 12.6 & 79.7 & 77.8 & 0.9 & 5.2 & 89.2 & 81.8 & 2.0 & 9.4 & 86.8 & 78.9 & 1.7 & 15.0 & 83.3 & 88.7 & 3.2 & 15.3 & 79.5 & 78.8 \\
      \rowcolor{ClosedColor} \small Gemini-2.5-pro-preview-05-06 & \faLock{} & 1.7 & 18.2 & 91.6 & \bf 90.6 & 2.4 & 11.2 & 87.6 & 78.6 & 2.1 & 11.9 & 93.7 & 82.4 & 0.9 & 3.8 & 90.6 & 75.0 & 1.7 & 6.0 & 96.0 & 70.8 & 0.0 & 8.3 & 85.0 & \bf 100.0 & 1.7 & 12.2 & 91.8 & \bf 86.1 \\
      \rowcolor{ClosedColor} \small GPT-4o-2024-11-20 & \faLock{} & 10.4 & 30.5 & 92.8 & 66.0 & 3.2 & 15.6 & 77.2 & 79.5 & 4.2 & 11.9 & 63.6 & 64.7 & 1.4 & 4.7 & 95.3 & 70.0 & 0.7 & 9.9 & 85.4 & \bf 92.5 & 40.0 & 76.7 & 80.0 & 47.8 & 6.8 & 21.3 & 86.8 & 67.9 \\
      \rowcolor{ClosedColor} \small GPT-4o-mini-2024-07-18 & \faLock{} & 7.6 & 19.8 & 88.2 & 61.7 & 0.4 & 7.6 & 80.0 & \bf 94.7 & 0.7 & 2.8 & 58.7 & 75.0 & 0.0 & 0.5 & 92.5 & \bf 100.0 & 0.5 & 2.7 & 89.6 & 81.8 & 26.7 & 43.3 & 90.0 & 38.3 & 4.3 & 11.8 & 85.7 & 63.2 \\
      \rowcolor{ClosedColor} \small GPT-4.1-nano-2025-04-14 & \faLock{} & 0.7 & 2.8 & 83.8 & 73.9 & 1.6 & 5.2 & 83.2 & 69.2 & 0.0 & 2.8 & 42.7 & \bf 100.0 & 0.0 & 0.0 & 95.3 & 0.0 & 0.2 & 0.7 & 83.4 & 66.7 & 1.7 & 5.0 & 81.7 & 66.0 & 0.6 & 2.4 & 81.7 & 73.9 \\
      \rowcolor{ClosedColor} \small GPT-4.1-mini-2025-04-14 & \faLock{} & 12.2 & 37.1 & 92.1 & 67.1 & 5.2 & 21.6 & 91.2 & 75.9 & 4.9 & 17.5 & 65.7 & 72.0 & 4.7 & 7.5 & 96.2 & 37.5 & 3.5 & 16.6 & 85.6 & 79.1 & \bf 60.0 & \bf 86.7 & 98.3 & 27.4 & 9.5 & 27.4 & 89.2 & 65.3 \\
      \rowcolor{ClosedColor} \small GPT-4.1-2025-04-14 & \faLock{} & 10.9 & 43.9 & 94.9 & 75.3 & 4.0 & 25.2 & 87.2 & 84.1 & 6.3 & 19.6 & 93.7 & 67.9 & 4.7 & 9.9 & 94.3 & 52.4 & 2.5 & 12.9 & 95.8 & 80.8 & 11.7 & 68.3 & 95.0 & 82.9 & 7.2 & 29.9 & 93.9 & 76.1 \\
      \rowcolor{ClosedColor} \small GPT-4.5-preview-2025-02-27 & \faLock{} & 10.4 & 45.7 & 97.1 & 77.3 & 6.0 & 26.0 & 88.8 & 76.9 & 6.3 & 25.2 & 95.8 & 75.0 & 4.7 & 9.4 & 92.5 & 50.0 & 1.7 & 10.4 & 95.5 & 83.3 & 5.0 & 58.3 & \bf 100.0 & 91.4 & 6.8 & 30.3 & 95.1 & 77.5 \\
      \rowcolor{ClosedColor} \small Grok-3-mini-fast & \faLock{} & 10.6 & 36.1 & 91.6 & 70.6 & 2.0 & 27.6 & 92.4 & 92.8 & 7.0 & 16.8 & 74.1 & 58.3 & 3.3 & 5.2 & 97.2 & 36.4 & 3.0 & 11.2 & 94.8 & 73.3 & 1.7 & 11.7 & 98.3 & 85.5 & 6.5 & 23.9 & 91.9 & 73.0 \\
      \rowcolor{ClosedColor} \small Grok-3-fast & \faLock{} & 16.6 & 42.8 & 97.0 & 61.3 & 8.4 & 27.6 & 96.0 & 69.6 & 5.6 & 16.8 & \bf 97.2 & 66.7 & 4.2 & 9.0 & 97.6 & 52.6 & 5.0 & 18.9 & 98.0 & 73.7 & 1.7 & 60.0 & \bf 100.0 & 97.2 & 10.3 & 30.5 & 97.2 & 66.1 \\
      \rowcolor{ClosedColor} \small Grok-3-mini & \faLock{} & 11.3 & 37.8 & 92.3 & 70.0 & 3.2 & 21.2 & 92.8 & 84.9 & 6.3 & 18.2 & 75.5 & 65.4 & 1.9 & 4.7 & 97.2 & 60.0 & 2.7 & 12.2 & 96.0 & 77.6 & 3.3 & 21.7 & 98.3 & 84.8 & 6.7 & 24.4 & 92.6 & 72.5 \\
      \rowcolor{ClosedColor} \small Grok-3 & \faLock{} & 15.7 & 45.0 & 98.0 & 65.0 & 6.8 & 26.4 & 95.2 & 74.2 & 6.3 & 15.4 & 95.1 & 59.1 & 2.8 & 7.5 & 97.6 & 62.5 & 5.2 & 22.1 & 97.5 & 76.4 & 0.0 & 56.7 & \bf 100.0 & \bf 100.0 & 9.6 & 31.6 & 97.4 & 69.5 \\
      \rowcolor{ClosedColor} \small o1-mini-2024-09-12 & \faLock{} & 9.8 & 28.7 & 88.7 & 66.0 & 1.2 & 15.2 & 84.4 & 92.1 & 6.3 & 15.4 & 61.5 & 59.1 & 1.9 & 3.8 & 90.1 & 50.0 & 2.2 & 7.4 & 92.6 & 70.0 & 20.0 & 45.0 & 83.3 & 55.6 & 64.9 & 19.1 & 86.9 & 67.5 \\
      \rowcolor{ClosedColor} \small o3-mini-2025-01-31 & \faLock{} & 18.2 & 50.5 & 98.2 & 64.0 & 3.2 & 30.0 & 95.6 & 89.3 & 8.4 & 28.7 & 72.7 & 70.7 & 4.7 & 6.6 & 95.8 & 28.6 & 3.5 & 20.6 & 97.8 & 83.1 & 25.0 & 71.7 & 96.7 & 65.1 & 11.0 & 35.5 & 95.5 & 69.0 \\
      \rowcolor{ClosedColor} \small o4-mini-2025-04-16 & \faLock{} & \bf 18.4 & \bf 51.8 & 95.7 & 64.5 & 5.2 & \bf 31.2 & 94.0 & 83.3 & \bf 11.9 & \bf 30.8 & 83.9 & 61.4 & \bf 5.2 & \bf 10.8 & 99.1 & 52.2 & 6.5 & \bf 22.3 & 93.5 & 71.1 & 45.0 & 58.3 & 98.3 & 22.8 & \bf 13.0 & \bf 36.8 & 94.6 & 64.7 \\
      \midrule
      \rowcolor{HeaderColor}\multicolumn{30}{c}{\textit{Open-Source LLMs}} \\
      \midrule
      \rowcolor{OpenColor} \small Qwen2.5-Coder-32B-Instruct & 32B & 6.2 & 19.4 & 82.0 & 67.9 & 3.6 & 14.8 & 74.0 & 75.7 & 3.5 & 4.9 & 34.3 & 28.6 & 0.5 & 1.9 & 79.7 & 75.0 & 2.0 & 5.7 & 78.7 & 65.2 & 0.0 & 23.3 & 28.3 & \bf 100.0 & 3.9 & 12.9 & 74.6 & 69.7 \\
      \rowcolor{OpenColor} \small Seed-Coder-8B-Instruct & 8B & 5.4 & 9.4 & 41.1 & 42.9 & 2.8 & 8.0 & 31.2 & 65.0 & 1.4 & 2.1 & 17.5 & 33.3 & 0.9 & 1.4 & 51.9 & 33.3 & 0.5 & 1.5 & 46.7 & 66.7 & 1.7 & 3.3 & 33.3 & 48.5 & 3.1 & 5.9 & 40.1 & 47.7 \\
      \rowcolor{OpenColor} \small DeepSeek-R1 & 37/671B & 8.9 & 35.4 & 96.3 & 74.8 & 7.2 & 22.4 & 97.2 & 67.9 & 5.6 & \bf 12.6 & 73.4 & 55.6 & 2.8 & \bf 7.5 & 96.2 & 62.5 & 3.0 & 15.1 & 96.5 & 80.3 & 0.0 & 30.0 & 100.0 & \bf 100.0 & 6.2 & 24.3 & 94.9 & 74.5 \\
      \rowcolor{OpenColor} \small DeepSeek-V3-250324 & 37/671B & 12.1 & \bf 40.4 & 98.5 & 70.1 & \bf 14.4 & 24.4 & 98.8 & 41.0 & \bf 6.3 & 11.2 & 93.7 & 43.8 & \bf 3.8 & 6.1 & 98.6 & 38.5 & 1.7 & 16.9 & 98.8 & 89.7 & 0.0 & 23.3 & 100.0 & \bf 100.0 & 8.4 & 26.6 & 98.3 & 68.4 \\
      \rowcolor{OpenColor} \small DeepSeek-V3 & 37/671B & 15.4 & 39.0 & 99.0 & 60.6 & 13.2 & \bf 24.8 & 99.2 & 46.8 & 5.6 & 11.9 & 91.6 & 52.9 & 3.3 & 6.1 & 99.5 & 46.2 & 2.0 & 15.4 & 98.8 & 87.1 & 0.0 & 28.3 & 98.3 & \bf 100.0 & 9.6 & 26.0 & 98.5 & 62.9 \\
      \rowcolor{OpenColor} \small QwQ-32B & 32B & 3.3 & 20.1 & 52.8 & \bf 83.6 & 0.8 & 2.8 & 50.8 & 71.4 & 0.0 & 8.4 & 33.6 & \bf 100.0 & 0.9 & 2.4 & 46.2 & 60.0 & 0.7 & 3.2 & 62.5 & 76.9 & 0.0 & 0.0 & 61.7 & 0.0 & 1.8 & 10.7 & 52.7 & \bf 83.2 \\
      \rowcolor{OpenColor} \small Qwen3-1.7B-Instruct Chat & 1.7B & 0.5 & 1.0 & 31.1 & 50.0 & 0.4 & 0.4 & 27.2 & 0.0 & 0.7 & 0.7 & 37.1 & 0.0 & 0.0 & 0.0 & 7.5 & 0.0 & 0.0 & 0.2 & 30.5 & \bf 100.0 & 0.0 & 1.7 & 65.0 & \bf 100.0 & 0.3 & 0.6 & 29.3 & 50.0 \\
      \rowcolor{OpenColor} \small Qwen3-1.7B-Instruct Think & 1.7B & 1.5 & 4.5 & 52.0 & 67.6 & 0.8 & 0.8 & 47.2 & 0.0 & 0.0 & 0.0 & 40.6 & 0.0 & 0.0 & 0.0 & 65.6 & 0.0 & 0.2 & 0.5 & 49.1 & 50.0 & 0.0 & 8.3 & 76.7 & \bf 100.0 & 0.8 & 2.4 & 52.2 & 67.4 \\
      \rowcolor{OpenColor} \small Qwen3-14B-Instruct Chat & 14B & 6.6 & 21.6 & 96.6 & 69.5 & 2.4 & 13.2 & 91.6 & 81.8 & 0.7 & 5.6 & 69.2 & 87.5 & 0.0 & 0.5 & 95.8 & \bf 100.0 & 2.0 & 6.2 & 93.8 & 68.0 & 0.0 & 25.0 & 83.3 & \bf 100.0 & 3.7 & 13.7 & 92.7 & 73.4 \\
      \rowcolor{OpenColor} \small Qwen3-14B-Instruct Think & 14B & 6.7 & 26.3 & 80.7 & 74.5 & 2.8 & 15.6 & 73.2 & 82.1 & 2.1 & 11.9 & 56.6 & 82.4 & 0.9 & 3.3 & 87.3 & 71.4 & 1.5 & 6.5 & 72.7 & 76.9 & 0.0 & 6.7 & 68.3 & \bf 100.0 & 3.9 & 16.4 & 76.5 & 76.4 \\
      \rowcolor{OpenColor} \small Qwen3-30B-A3B-Instruct Think & 3/30B & 7.2 & 23.5 & 75.4 & 69.4 & 2.0 & 7.6 & 72.4 & 73.7 & 1.4 & 9.1 & 57.3 & 84.6 & 1.4 & 1.9 & 57.1 & 25.0 & 1.0 & 6.7 & 79.2 & 85.2 & 11.7 & 26.7 & 81.7 & 56.2 & 4.2 & 14.4 & 72.6 & 70.6 \\
      \rowcolor{OpenColor} \small Qwen3-30B-A3B-Instruct Chat & 3/30B & 11.1 & 27.7 & 96.7 & 59.9 & 2.0 & 10.4 & 94.8 & 80.8 & 3.5 & 7.0 & 79.0 & 50.0 & 0.0 & 0.5 & 91.0 & \bf 100.0 & 2.5 & 6.5 & 94.8 & 61.5 & 5.0 & 25.0 & 95.0 & 80.0 & 6.0 & 16.2 & 94.0 & 62.6 \\
      \rowcolor{OpenColor} \small Qwen3-32B-Instruct Think & 32B & 6.8 & 29.0 & 82.7 & 76.5 & 4.8 & 16.4 & 82.8 & 70.7 & 2.8 & 10.5 & 58.0 & 73.3 & 1.9 & 3.3 & 72.2 & 42.9 & 2.2 & 8.7 & 83.9 & 74.3 & 13.3 & 23.3 & 75.0 & 42.9 & 4.9 & 18.5 & 79.7 & 73.4 \\
      \rowcolor{OpenColor} \small Qwen3-32B-Instruct Chat & 32B & 7.4 & 23.8 & 90.5 & 68.7 & 5.6 & 14.8 & 92.0 & 62.2 & 4.9 & 7.7 & 84.6 & 36.4 & 0.5 & 2.4 & 93.9 & 80.0 & 1.7 & 7.4 & 95.0 & 76.7 & 13.3 & 36.7 & 95.0 & 63.8 & 5.2 & 15.9 & 91.7 & 67.3 \\
      \rowcolor{OpenColor} \small Qwen3-4B-Instruct Chat & 4B & 3.9 & 11.2 & 87.9 & 65.2 & 0.8 & 1.2 & 39.6 & 33.3 & 1.4 & 1.4 & 62.9 & 0.0 & 0.5 & 0.5 & 84.0 & 0.0 & 0.7 & 1.2 & 40.7 & 40.0 & 0.0 & 3.3 & 45.0 & \bf 100.0 & 2.1 & 5.6 & 67.7 & 61.9 \\
      \rowcolor{OpenColor} \small Qwen3-4B-Instruct Think & 4B & 4.4 & 13.8 & 72.0 & 68.1 & 1.6 & 3.6 & 65.2 & 55.6 & 0.0 & 2.1 & 46.9 & \bf 100.0 & 0.0 & 0.9 & 66.0 & \bf 100.0 & 0.0 & 1.7 & 60.5 & \bf 100.0 & 0.0 & 11.7 & 48.3 & \bf 100.0 & 2.1 & 7.5 & 65.3 & 71.6 \\
      \rowcolor{OpenColor} \small Qwen3-8B-Instruct Chat & 8B & 6.5 & 17.3 & 93.0 & 62.7 & 8.8 & 14.8 & 95.6 & 40.5 & 0.7 & 2.1 & 76.9 & 66.7 & 0.0 & 0.9 & 94.8 & \bf 100.0 & 1.0 & 4.0 & 92.6 & 75.0 & \bf 25.0 & 46.7 & 83.3 & 46.5 & 5.0 & 12.1 & 91.9 & 58.3 \\
      \rowcolor{OpenColor} \small Qwen3-8B-Instruct Think & 8B & 5.7 & 19.9 & 70.9 & 71.2 & 2.0 & 6.0 & 77.2 & 66.7 & 0.7 & 6.3 & 53.1 & 88.9 & 0.0 & 0.9 & 65.1 & \bf 100.0 & 1.7 & 4.5 & 74.2 & 61.1 & 1.7 & 10.0 & 43.3 & 83.0 & 3.2 & 11.3 & 69.5 & 71.4 \\
      \rowcolor{OpenColor} \small Qwen3-0.6B-Instruct Chat & 0.6B & 0.2 & 0.2 & 17.9 & 0.0 & 0.4 & 0.4 & 11.2 & 0.0 & 0.0 & 0.0 & 21.0 & 0.0 & 0.0 & 0.0 & 11.8 & 0.0 & 0.0 & 0.0 & 10.7 & 0.0 & 0.0 & 0.0 & 10.0 & 0.0 & 0.2 & 0.2 & 14.8 & 0.0 \\
      \rowcolor{OpenColor} \small Qwen3-0.6B-Instruct Think & 0.6B & 0.2 & 0.2 & 12.3 & 0.0 & 0.8 & 0.8 & 4.0 & 0.0 & 0.0 & 0.0 & 13.3 & 0.0 & 0.0 & 0.0 & 16.0 & 0.0 & 0.0 & 0.0 & 8.7 & 0.0 & 0.0 & 0.0 & 3.3 & 0.0 & 0.2 & 0.2 & 10.6 & 0.0 \\
      \rowcolor{OpenColor} \small OpenCoder-8B-Instruct & 8B & 0.2 & 0.2 & \bf 100.0 & 0.0 & 0.8 & 0.8 & \bf 100.0 & 0.0 & 0.0 & 0.0 & \bf 100.0 & 0.0 & 0.0 & 0.0 & \bf 100.0 & 0.0 & 0.0 & 0.0 & \bf 100.0 & 0.0 & 0.0 & 0.0 & \bf 100.0 & 0.0 & 0.2 & 0.2 & \bf 100.0 & 0.0 \\
      \rowcolor{OpenColor} \small DeepSeek-R1-Distill-Llama-8B & 8B & 1.0 & 2.8 & 72.2 & 65.2 & 0.0 & 0.4 & 74.8 & 100.0 & 0.0 & 0.0 & 24.5 & 0.0 & 0.0 & 0.0 & 70.8 & 0.0 & 0.2 & 0.2 & 52.9 & 0.0 & 28.3 & 30.0 & 78.3 & 5.6 & 1.4 & 2.3 & 64.8 & 39.5 \\
      \rowcolor{OpenColor} \small DeepSeek-R1-Distill-Qwen-14B & 14B & 4.1 & 11.2 & 91.7 & 63.0 & 0.4 & 2.0 & 91.2 & 80.0 & 0.7 & 2.1 & 39.2 & 66.7 & 0.0 & 0.0 & 94.3 & 0.0 & 0.2 & 0.7 & 81.6 & 66.7 & 1.7 & 5.0 & 91.7 & 66.7 & 2.0 & 5.6 & 85.8 & 64.2 \\
      \rowcolor{OpenColor} \small Llama3.1-8B-Instruct & 8B & 0.2 & 0.2 & 2.6 & 0.0 & 0.8 & 0.8 & 4.8 & 0.0 & 0.0 & 0.0 & 3.5 & 0.0 & 0.0 & 0.0 & 5.2 & 0.0 & 0.0 & 0.0 & 4.2 & 0.0 & 0.0 & 0.0 & 6.7 & 0.0 & 0.2 & 0.2 & 3.7 & 0.0 \\
      \midrule
      \rowcolor{OpenColor} \small \coder & 32B & \bf 15.7 & 37.3 & 96.2 & 57.8 & 12.0 & 22.0 & 92.0 & 45.5 & 4.9 & 10.5 & 81.8 & 53.3 & 2.8 & 4.7 & 96.7 & 40.0 & \bf 5.2 & \bf 18.1 & 95.8 & 71.2 & 10.0 & \bf 85.0 & 100.0 & 88.2 & \bf 10.5 & \bf 27.0 & 94.7 & 61.0 \\
      \bottomrule
    \end{tabular}%
  }%
  \caption{Main Results for `\fix`.}
  \label{tab:results_diff}
  \vspace{-20pt}
\end{table*}

\section{Experiments}
\subsection{Implementation Details}
We fine-tune Qwen2.5-Coder-32B on 128 NVIDIA H800-80GB GPUs\footnote{\url{https://github.com/QwenLM/Qwen2.5-Coder/tree/main/finetuning/sft}} using the \instruction{} dataset and a set of roughly 500 K repositories, employing Adam with a cosine-decay scheduler: after 100 warm-up steps the learning rate peaks at 5e-5, a global batch size of 1,024, and inputs truncated to 32,768 tokens. In a second stage, we further train \coder{} on 150,000 simulated multi-turn dialogue trajectories—covering code synthesis, iterative debugging, and style standardization—to sharpen its practical generative and reasoning skills in realistic coding scenarios.

\subsection{Evaluation Metrics}
Inspired by the aider polyglot benchmark setting, we evaluate LLMs with four key metrics:

\noindent\textbf{Pass@1.} Measures the proportion of coding tasks an LLM completes correctly on its first attempt, as verified by test cases \cite{codegeex}, directly reflecting the one-shot coding accuracy of LLMs.

\noindent\textbf{Pass@2.} After a failed attempt, LLMs can view their previous code and error messages before trying again, capturing the capacity of LLMs to improve via immediate feedback.

\noindent\textbf{Fix Weight (FW).} Defined as $\mathrm{FW} \;=\;\frac{\mathrm{Pass@2} - \mathrm{Pass@1}}{\mathrm{Pass@2}}$, it represents the fraction of successful second-attempt fixes among all second-attempt successes, emphasizing the relative contribution of LLM through error diagnosis and correction.

\noindent\textbf{Well Format (WF).} Measures the percentage of tasks where the LLM strictly follows the edit format specified in the system prompt, quantifying format compliance under constrained instructions.

\subsection{Evaluation Edit Format}

\autoref{fig:edit_format} shows two different ``edit formats'' when evaluating different LLMs. The ``\new'' format is the easiest for an LLM to use, but it uses a lot of tokens and may limit how large a file can be edited. Models which can use one of the ``\fix'' formats are much more efficient, using far fewer tokens. Models that use a diff-like format are able to edit larger files with less cost and without hitting token limits. For fair comparison, we will show the scores in ``\new'' and ``\fix'' format to fully evaluate all LLMs.

%%%%%%%%%%%%%%%%%%%%%%%%%%%%%%%%%%%%%%%%%%%%%%%%%%%%%%%%%%%%%%%%%%%%%%
\begin{figure*}[!t]
\centering
\includegraphics[width=1.0\textwidth]{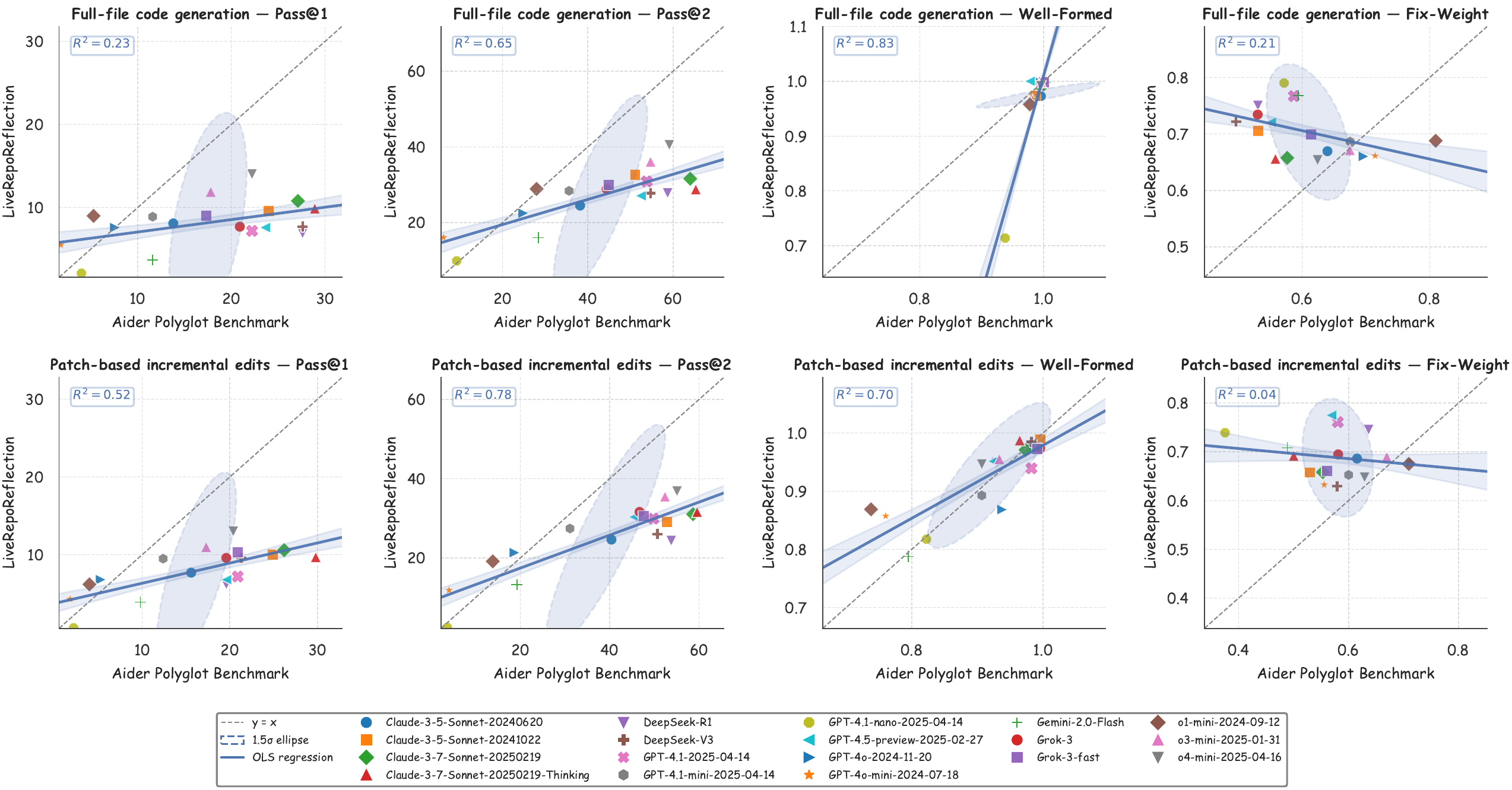}
\caption{Performance comparison between \bench{} and the Aider Polyglot Benchmark over multiple large language models. Scatter points show individual model scores under full‐file code generation and patch‐based incremental edits across four metrics (Pass@1, Pass@2, Well-Formed, Fix-Weight). The dashed diagonal line denotes $y=x$, ellipses represent 1.5$\sigma$ confidence regions, and solid lines are ordinary least squares fits annotated with their $R^2$ values.}
\label{fig:aider_polyglot_comparison}
\vspace{-10pt}
\end{figure*}
%%%%%%%%%%%%%%%%%%%%%%%%%%%%%%%%%%%%%%%%%%%%%%%%%%%%%%%%%%%%%%%%%%%%%%

\subsection{Code LLMs}
We evaluate 40+ LLMs, including GPT-4.1/GPT-4.5~\citep{gpt45}, Claude-3.5/3.7~\citep{claude2,claude37}, o1-mini/o4-mini~\citep{o1_mini}, Gemini~\citep{gemini}, Qwen2.5-Coder/Qwen3/Qwq-32B~\citep{qwen25coder,qwen3,qwq-32b}, Grok~\citep{grok3}, OpenCoder~\citep{opencoder}, DeepSeek-R1/V3~\citep{deepseek_r1,deepseek_v3} and Doubao-1-5-Thinking-pro-m-250415~\citep{doubao-thinking}. And to compare all LLMs fairly, we use a temperature $0$, max\_output\_tokens $8192$ and no additional custom parameters.

\subsection{Main Result}

\autoref{tab:results_whole} and \autoref{tab:results_diff} show that the results under both \fix~and \new~formats show that leading closed-source models consistently achieve the highest one-shot and post-feedback accuracies, while open‐source models trail behind but display similar relative gains when allowed a second attempt. All systems find Python tasks easiest and struggle most with C++ and Rust. Nearly every model exceeds ninety percent format-compliance, and fix-weight values around sixty to eighty percent demonstrate that test feedback reliably drives improvements. Incremental patch edits narrow the gap between first and second passes compared with full-file generation. Our \coder{} clearly outperforms its Qwen2.5‐Coder base yet still falls short of the top closed-source performers, highlighting room for further advances in multi‐file code reflection.

%%%%%%%%%%%%%%%%%%%%%%%%%%%%%%%%%%%%%%%%%%%%%%%%%%%%%%%%%%%%%%%%%%%%%%
\begin{figure*}[t!]
\centering
\includegraphics[width=1.0\textwidth]{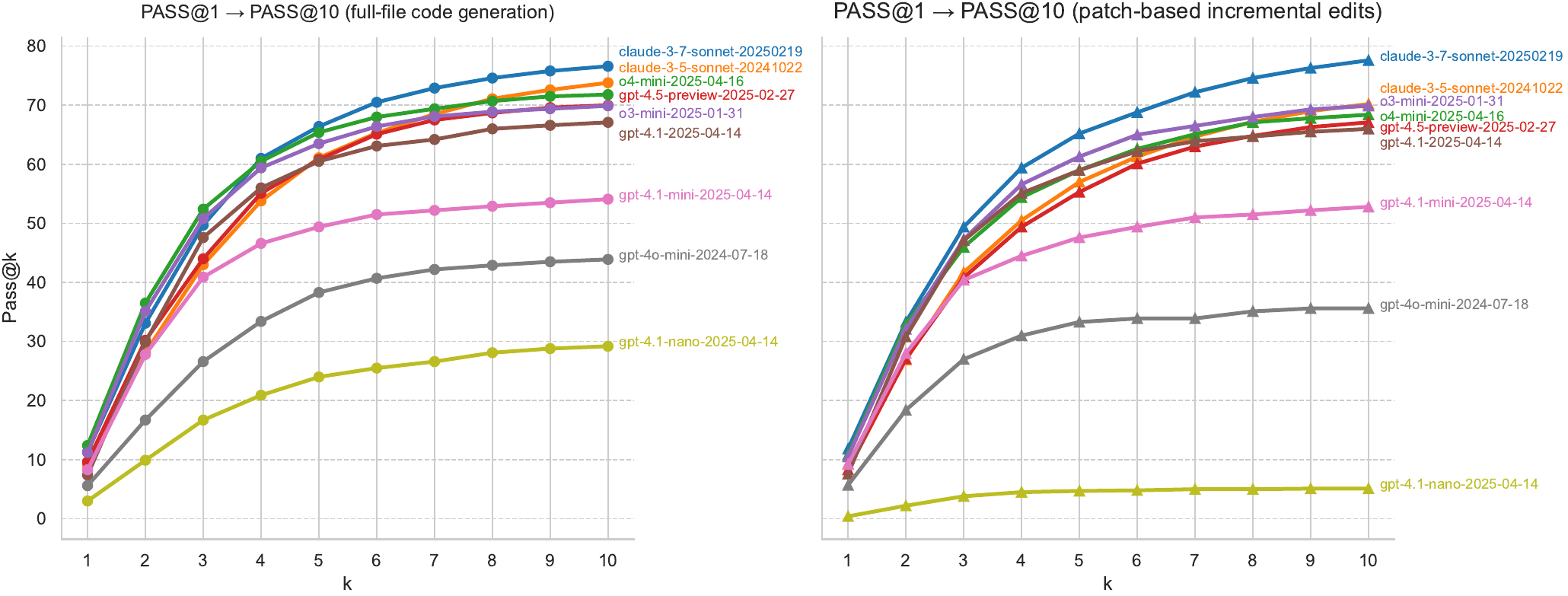}
\caption{Pass@k (k=1–10) curves for nine LLMs under full‐file code generation (left) and patch‐based incremental editing (right). All models exhibit strictly increasing performance with diminishing marginal gains beyond k=2; patch‐based editing yields more uniform improvements across samples, whereas full‐file generation achieves higher absolute pass rates.}
\label{fig:pass1_to_10}
\end{figure*}
%%%%%%%%%%%%%%%%%%%%%%%%%%%%%%%%%%%%%%%%%%%%%%%%%%%%%%%%%%%%%%%%%%%%%%

\section{Analysis}

\subsection{Performance Comparison between \bench{} and Aider Polyglot Benchmark}

\autoref{fig:aider_polyglot_comparison} reveals that while pass@1 and pass@2 scores on \bench{} correlate moderately with the older aider polyglot benchmark ($R^2\approx0.65$ for pass@1 and $0.78$ for pass@2), nearly all points fall below the $y=x$ line, indicating consistently lower absolute pass rates on \bench{} and thus greater task difficulty. well-formedness remains highly consistent between the two datasets ($R^2\approx0.83$ full-file, $0.70$ diff-based), showing that syntactic and structural constraints are enforced similarly. In stark contrast, fix-weight exhibits very weak alignment ($R^2\approx0.21$ and $0.04$), underscoring divergent repair performance and demonstrating that \bench{} more reliably challenges and discriminates model debugging capabilities. Moreover, because the aider polyglot benchmark has been publicly available since mid-2024 and may have been subject to overfitting by recent model releases, our freshly curated \bench{}, compiled from never-before-seen repositories through May 2025, provides a more robust and less biased evaluation of real-world code generation and repair.

\subsection{Pass@k Curves for Two Editing Formats}

We evaluate 9 representative LLMs under two edit formats (\new~and~\fix) by measuring pass@k ($k = 1->10$). \autoref{fig:pass1_to_10} illustrates pass@k of each LLM curve is strictly increasing yet exhibits pronounced diminishing returns: the largest marginal gain occurs at k=1→2, and thereafter each additional sample yields progressively smaller improvements. Although \new~attains higher absolute pass rates, \fix~produces notably more uniform gains across k, suggesting that iterative feedback helps smaller improvements stack more evenly. Some LLMs  (e.g. o3-mini, claude-3-5-sonnet) enjoy larger early gains but plateau by k≈5–7, and other LLMs (e.g. o4-mini, gpt-4.1) start with high pass@1 and maintain smoother, lower-amplitude increments. The overall trends of all LLMs are consistent. As the number of iterations k increases and the feedback information increases, pass@k gradually increases, and the rate of increase gradually slows down. In addition, weaker LLMs (such as gpt4.1-nano-2025-04-14 and gpt4o-mini-2024-07-18) perform worse than \new~in \fix, but there is no significant difference in stronger LLMs.

\section{Related Work}
\paragraph{Code Large Language Models and Reflection.} Pre‐training advances in NLP have improved code understanding in models like CodeBERT \citep{CodeBERT} and CodeT5 \citep{CodeT5}, and spurred code‐specific LLMs such as CodeGen \citep{codegen}, Code Llama \citep{code_llama} and StarCoder \citep{starcoder}. Leveraging instruction tuning \citep{instructGPT,self_instructions}, methods like Evol-Instruct \citep{wizardcoder} and OSS-Instruct \citep{magicoder} generate multilingual code instructions evaluated by benchmarks \citep{multiple,mceval}. Repository-based tests (SWEBench \citep{swebench}, Aider-polyglot \citep{aider}, ExecRepoBench \citep{execrepobench}) measure real-world code tasks and reflection, with DeepSeek-R1 \citep{deepseek_r1} and ReflectionCoder \citep{reflectioncoder} demonstrating advanced reflection capabilities.

\section{Conclusion}

This paper proposes \bench{}, a high-difficulty code reflection benchmark for multi-file repository scenarios. By combining automated pipelines with manual verification, we built a total of 1,888 strictly screened test cases covering six programming languages. Based on multi-source high-quality corpora and multi-round dialogue generation strategies, we built the \instruction{} instruction set and trained \coder{}, achieving significant performance improvements. Experimental results show that \bench{} can truly and effectively measure the model's reflection and repair capabilities in cross-file dependency and iterative repair scenarios, providing a solid foundation for subsequent research.

\clearpage
\section{Limitations}
\label{sec:limitations}
We acknowledge the following limitations of this study: (1) The evaluation in repository-level multilingual scenarios are not fully explored. (2) The code completion model \coder{} is mainly supervised fine-tuned on the 7B open-source base LLMs. In the future, we will try the (3) The fine-tuned model can be further improved using RLHF for better user experience, such as DPO.

\section*{Ethics Statement}
This research adheres to ethical guidelines for AI development. We aim to enhance the capabilities of large language models (LLMs) while acknowledging potential risks such as bias, misuse, and privacy concerns. To mitigate these, we advocate for transparency, rigorous bias testing, robust security measures, and human oversight in AI applications. Our goal is to contribute positively to the field and to encourage responsible AI development and deployment.
\bibliography{custom}
\bibliographystyle{acl_natbib}

\appendix

\onecolumn

\clearpage

\section{Related Work}\label{related_work}

\paragraph{Code Large Language Model.}
Leveraging advancements in NLP, pre-training techniques have significantly bolstered code understanding and synthesis in models like CodeBERT \citep{CodeBERT} and CodeT5 \citep{CodeT5}, leading to the adoption of NLP-inspired architectures and objectives for tasks such as code generation, infilling, summarization, refinement, and translation \citep{CodeXGLUE,CodeTransOcean,refine_gpt,chat_unitest}. The emergence of code-specific large language models (LLMs) \citep{starcoder,code_llama,guo2024deepseekcoder,codearena,execrepobench}, exemplified by CodeGen \citep{codegen} and Code Llama \citep{code_llama}, demonstrates foundational competence in code understanding and generation. Inspired by multi-agent collaboration \citep{multi_agents_survey,autonomous_agents_survey}, the concept of language-specific agents is introduced to create multilingual instruction datasets, building upon the success of instruction tuning \citep{instructGPT,llama_adapter,self_instructions} and innovations like code Evol-Instruct \citep{wizardcoder} and the utilization of real-world code in OSS-Instruct~\citep{magicoder} and CodeOcean~\citep{wavecoder} to enhance instruction data quality and realism, with multilingual benchmarks \citep{multiple,mceval,mdeval,fullstack,bigcodebench} assessing these models' capabilities.

\paragraph{Code Reflection}

For code intelligence, repository-based code benchmarks such as SWEBench~\citep{swebench,swe_lancer}, Aider-polyglot~\citep{aider}, and ExecRepoBench~\citep{execrepobench} evaluate the abilities of addressing real-world software engineering tasks, code reflection, and repository-based code completion. DeepSeek-R1~\citep{deepseek_r1} demonstrates capabilities such as reflection capabilities in long CoTs, marking a significant milestone for the research community. Reflectioncoder~\citep{reflectioncoder}

\section{Polyglot Repository Code File Structure Examples}\label{appendix_polyglot_repository_code_file_structure}

% Polyglot Repository Code File Structure Examples of \bench{} and instruction corpus of \instruction{}

\begin{figure*}[h]
\begin{center}
    \includegraphics[width=1.0\textwidth]{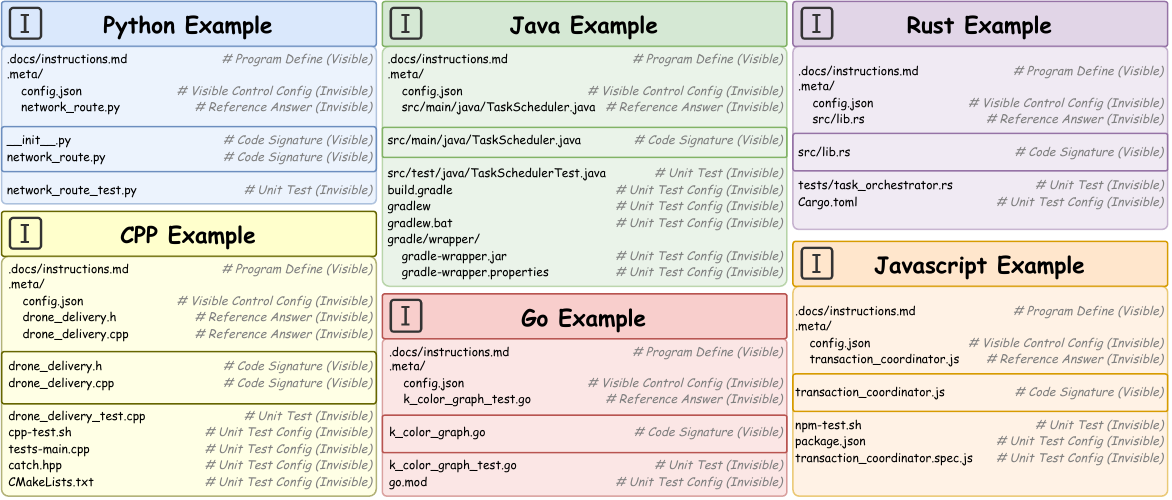}
    \caption{Polyglot repository code file structure examples of \bench{} and instruction corpus of \instruction{}. Visibility (``Visiable'' or ``Invisiable'') indicates whether the model is allowed to obtain and modify the file content during evaluation. The repository code file structure consists of five parts: 1) problem definition, 2) reference answer, 3) code signature, 4) unit test, 5) unit test environment support file.}
    \label{fig:file_structure_examples}
\end{center}
\end{figure*}

We provide polyglot repository code file structure examples of \bench{} and instruction corpus of \instruction{} in \autoref{fig:file_structure_examples}.

\section{Language Distribution Comparison}\label{language_distribution_comparison}
Table~\ref{tab:language_distribution} compares the number of test cases per programming language in \bench{} versus the Aider Polyglot Benchmark. \bench{} contains over eight times as many test cases overall (1,888 vs. 225) and expands coverage in every language. Python, as a dominant scripting language, grows from 34 to 820 cases, while compiled languages like Go (39→403), Java (47→250), C++ (26→143), and Rust (30→212) also see substantial increases. JavaScript remains smaller in absolute terms but still benefits from a slight boost (49→60). This balanced, large-scale distribution across both scripting and system languages ensures that models are evaluated on a wide spectrum of real-world coding tasks and paradigms.

\begin{table}[h]
  \centering
  \small
  \caption{Language Distribution Comparison}
  \label{tab:language_distribution}
  \begin{tabular}{lrr}
    \toprule
    Language      & \bench{} & Aider Polyglot Benchmark \\
    \midrule
    C++           & 143                        & 26                   \\
    Go            & 403                        & 39                   \\
    Java          & 250                        & 47                   \\
    JavaScript    & 60                         & 49                   \\
    Python        & 820                        & 34                   \\
    Rust          & 212                        & 30                   \\
    \midrule
    Total         & 1,888                      & 225                  \\
    \bottomrule
  \end{tabular}
\end{table}

\section{Prompts for Our Pipeline, \bench{} and \instruction{}}\label{prompt}

We provided data generation stage prompts for our pipeline, \bench{} and \instruction{} in \autoref{prompt_figure_1}, \autoref{prompt_figure_2}, \autoref{prompt_figure_10}, \autoref{prompt_figure_3}, \autoref{prompt_figure_4}, \autoref{prompt_figure_5}, \autoref{prompt_figure_6} and \autoref{prompt_figure_7}.

\boxfigure{}{
Act as a high-level programming competition question setter and take requests for generating a new code problem.\\\\1. Make sure your code problem concise but complete.\\2. Make sure your code problem difficult and challenging.
}
{System Prompt for Our Pipeline.}
{prompt_figure_1}

\boxfigure{}{
When Creating files, maintain a consistent folder structure as shown in the examples by providing the appropriate path/to/filename and adhere to the following format:
\\\\
path/to/filename\\
```\\
// entire code or file content ...\\
```
\\\\
- The first line should contain *only* the appropriate path/to/filename, without any additional markup, punctuation, comments, or other elements.\\
- The second line should start with three backticks (```)\\
- ... include the complete content of the file ...\\
- The last line should end with three closing backticks (```)
\\\\
Please ensure that you *never* skip, omit, or abbreviate content using ellipsis (...) or by adding comments like ``... rest of code...''. Use only standard libraries in your code.
}
{Format Reminder Prompt for Our Pipeline.}
{prompt_figure_2}

\boxfigure{}{
Here are some question examples to give you inspiration:
\\\\
\#\# Question Example \{index\_placeholder\}:
\\\\
\#\#\# Project Name
\\\\
\{project\_name\}
\\\\
\#\#\# Question Description
\\\\
\{question\_description\}
\\\\
\#\#\# Answer File With Dependencies
\\\\
\{answer\}
\\\\
\#\#\# Unit Test File
\\\\
\{unit\_test\}
}
{Sample Data Template Prompt for Our Pipeline.}
{prompt_figure_10}

\boxfigure{}{
Please generate a challenging and sophisticated \{language\} coding problem. Consider incorporating these elements to increase complexity from question example inspiration:
\\\\
- Advanced data structures (trees, graphs, heaps)\\
- Multiple edge cases and constraints\\
- Optimization requirements\\
- Real-world practical scenarios\\
- System design aspects\\
- Algorithmic efficiency requirements\\
- Multiple valid approaches with different trade-offs
\\\\
Make sure the difficulty is as high as possible, similar to the leetcode Hard level.
\\\\
Please **only** describe the question clearly but challenge the solver with interesting constraints and requirements. Do not include any project name, code signature, answer, unit test or any other things at this stage.
\\\\
\{sample\_data\_str\}
\\\\
Now, begin!
}
{Coding Program Definition Prompt for Our Pipeline.}
{prompt_figure_3}

\boxfigure{}{
Now, present the name of this \{language\} problem with snake case and keep it short and concise by using 1-3 words, like ``hello\_world''. Please generate the name in the following format: 
\\\\
```
project\_name
```
\\\\
Please **only** generate the name in the above format. Do not include any question description, code signature, answer, unit test or any other things at this stage. Now, begin!
}
{Coding Program Topic for Our Pipeline.}
{prompt_figure_4}

\boxfigure{}{
Please supply a comprehensive \{language\} unit test for this question. DO NOT include any answer or any other things at this stage.
\\\\
\{format\_reminder\}
\\\\
Attention to follow and implement the `\{project\_name\}` project structure, each file should replace in `\{project\_name\}` folder and have similar filepath to ensure the unit test can be run successfully.
\\\\
Now, begin! \{end\_suffix\}
}
{Unit Test Prompt for Our Pipeline.}
{prompt_figure_5}

\boxfigure{}{
Please supply a comprehensive \{language\} answer and necessary dependencies for this question.
\\\\
\{format\_reminder\}
\\\\
Attention to follow and implement the `\{project\_name\}` project structure, each file should replace in `\{project\_name\}` folder and have similar filepath to ensure the answer can be run successfully.
\\\\
Now, begin! \{end\_suffix\}
}
{Reference Answer Prompt for Our Pipeline.}
{prompt_figure_6}

\boxfigure{}{
1. Attention Our JavaScript Environment is `Node.js v16.20.2`.
\\
2. Attention Our Rust Environment is `rustc 1.75.0 (82e1608df 2023-12-21) (built from a source tarball)`, only support rust edition <= 2021.
}
{Optional End Suffix Prompt Examples for Our Pipeline.}
{prompt_figure_7}

\section{Unit Test Running Command Setup}

We provided unit test commands for our pipeline, \bench{} and \instruction{} in \autoref{fig:unit_test_command_setup}.

\boxfigure{}{
``python'': ``pytest''\\
``rust'': ``cargo test -- --include-ignored''\\
``go'': ``go test ./...''\\
``javascript'': ``./npm-test.sh''\\
``cpp'': ``./cpp-test.sh''\\
``java'': ``./gradlew test --no-daemon''
}
{Unit Test Command Setup.}
{fig:unit_test_command_setup}

\end{CJK*}
\end{document}